\documentclass[conference]{IEEEtran}

\usepackage{cite}
\usepackage{amsmath,amssymb,amsfonts}

\usepackage{graphicx}
\usepackage{subcaption}
\usepackage{textcomp}
\usepackage{xcolor}
\usepackage{xspace}
\usepackage{subcaption}
\usepackage{multirow}
\usepackage{booktabs} % For nicer table layout
\usepackage{algorithm}
\usepackage{algpseudocodex}
\algrenewcommand\algorithmicrequire{\textbf{Input:}}
\algrenewcommand\algorithmicensure{\textbf{Output:}}
\usepackage{float}
\usepackage{microtype}
\usepackage{layouts}
\usepackage{pgfplots}
\usepackage{pgfplots}
\usepackage{tabularx}
\usetikzlibrary{intersections}
\usepgfplotslibrary{fillbetween}

\usepackage[hidelinks]{hyperref}

\usepackage[capitalise]{cleveref}

\usepackage{ulem}
\usepackage{tikz}
\usetikzlibrary{shapes.geometric, arrows, positioning}
\usetikzlibrary{shadows,arrows,positioning}
\pgfdeclarelayer{background}
\pgfdeclarelayer{foreground}
\pgfsetlayers{background,main,foreground}
\usepackage{pgfplotstable}
\usepackage{scalefnt}
\usetikzlibrary{patterns}

\newcommand\extrap{\mbox{Extra-P}\xspace}
\newcommand\scorep{\mbox{Score-P}\xspace}

\AtBeginDocument{%
  \providecommand\BibTeX{{%
    \normalfont B\kern-0.5em{\scshape i\kern-0.25em b}\kern-0.8em\TeX}}}

\newif\ifourcomments
\ourcommentstrue 
\ifourcomments

\newcommand{\RemarkGustavo}[1]{\textcolor{red}{[Gustavo: #1]}}
\newcommand{\RemarkGregor}[1]{\textcolor{orange}{[Gregor: #1]}}
\newcommand{\RemarkAhmad}[1]{\textcolor{cyan}{[Ahmad: #1]}}
\newcommand{\RemarkFelix}[1]{\textcolor{magenta}{[Felix: #1]}}

\newcommand{\RemarkAlex}[1]{\textcolor{violet}{[Alexander: #1]}}
\newcommand{\lex}[1]{\textcolor{cyan}{[Lex: #1]}}
\else

\newcommand{\RemarkGustavo}[1]{}
\newcommand{\RemarkGregor}[1]{}
\newcommand{\RemarkAhmad}[1]{}
\newcommand{\RemarkFelix}[1]{}
\newcommand{\RemarkAlex}[1]{}
\newcommand{\lex}[1]{}
\fi

\begin{document}

\title{Denoising Application Performance Models with Noise-Resilient Priors}
%\author{ Anonymous Author(s)}

% ---------------
% Authors
% ---------------
\author{%
Gustavo de Morais\textsuperscript{1}, 
Alexander Geiß\textsuperscript{1}, 
Alexandru Calotoiu\textsuperscript{2}, 
Gregor Corbin\textsuperscript{3}, 
Ahmad Tarraf\textsuperscript{1}, \\
Torsten Hoefler\textsuperscript{2}, 
Bernd Mohr\textsuperscript{3}, 
Felix Wolf\textsuperscript{1} \\[1em]
\textsuperscript{1} Technical University of Darmstadt, Department of Computer Science, Darmstadt, Germany \\ 
\textsuperscript{2} ETH Zurich, Department of Computer Science, Zurich, Switzerland \\ 
\textsuperscript{3} Forschungszentrum J\"ulich GmbH, J\"ulich Supercomputing Centre, J\"ulich, Germany
\\
\texttt{\{gustavo.morais,alexander.geiss1,ahmad.tarraf,felix.wolf\}@tu-darmstadt.de} \\ 
\texttt{\{alexandru.calotoiu,htor\}@inf.ethz.ch} \\ 
\texttt{\{g.corbin,b.mohr\}@fz-juelich.de} 
}

\maketitle

\begin{abstract}
As parallel codes are scaled to larger computing systems, performance models play a crucial role in identifying potential bottlenecks. However, constructing these models analytically is often challenging. Empirical models based on performance measurements provide a practical alternative, but measurements on high-performance computing (HPC) systems are frequently affected by noise, which can lead to misleading predictions. To mitigate the impact of noise, we introduce application-specific dynamic priors into the modeling process. These priors are derived from noise-resilient measurements of computational effort, combined with domain knowledge about common algorithms used in communication routines. By incorporating these priors, we effectively constrain the model’s search space, eliminating complexity classes that capture noise rather than true performance characteristics. This approach keeps the models closely aligned with theoretical expectations and substantially enhances their predictive accuracy. Moreover, it reduces experimental overhead by cutting the number of repeated measurements by half. 
\end{abstract}


% Make the title area
\maketitle

%\IEEEpeerreviewmaketitle

\section{Introduction}

Performance models are essential for identifying bottlenecks in the early stages of HPC applications. These models express a performance metric of interest, such as execution time or energy consumption, as a mathematical function of one or more execution parameters, like input size or number of processors. 
Although often based on simplifying assumptions, performance models offer valuable insight at a small cost of evaluating an arithmetic expression~\cite{torsten2011}.  For example, a performance model such as $t(p) = c \cdot log(p)$, where $t$ is the runtime, $c$ is a constant coefficient, and $p$ is the number of MPI ranks, can tell how the code scales when changing the number of processes.
However, deriving such models analytically from the code is still so laborious that too many application developers shy away from the effort. Empirical performance modeling offers a practical alternative to address this challenge: instead of manually constructing the model, it is learned from performance measurements. However, this approach must contend with performance variability---the difference between execution times across repeated runs of an application in the same execution environment~\cite{patki_ea:sc19}. Particularly as a consequence of modern network topologies, the runtime on HPC systems can vary by up to $70\%$~\cite{chunduri2017run}.   

\begin{figure} 
\centerline{\includegraphics[width=\linewidth]{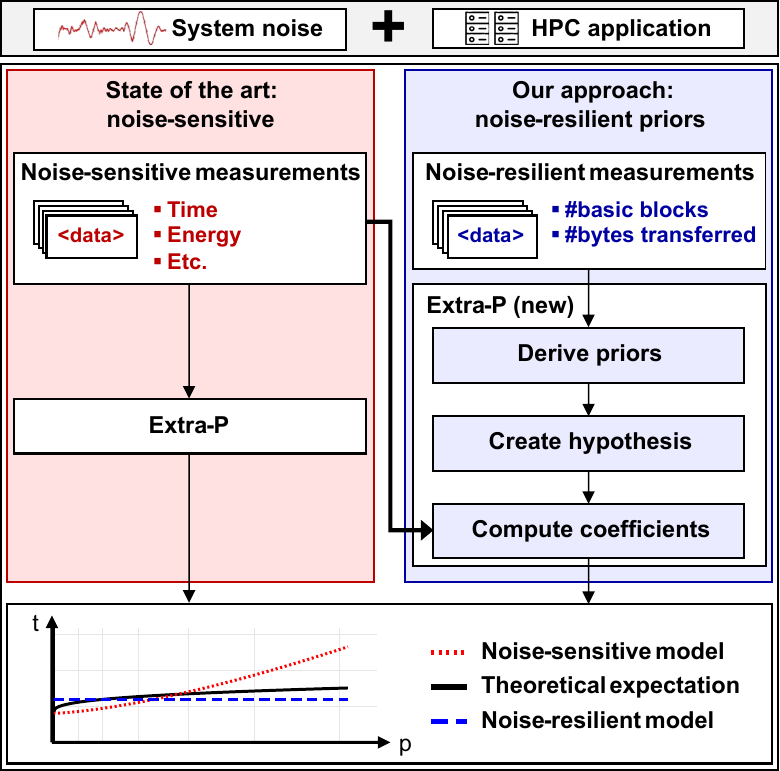}}
\caption{Noise-resilient empirical performance modeling with Extra-P. The noise-resilient model aligns with the theoretical expectation more closely.}
\label{fig:method}
\end{figure}

Performance variability is caused by a combination of node-level effects, such as sudden OS activities, dynamic frequency scaling, manufacturing variability, or shared-cache contention, and system-level effects, such as network and file-system congestion in the presence of concurrently running jobs. Since these events seem to occur randomly, we refer to them in this paper as noise.  The conventional method to mitigate noise involves repeating measurements and selecting a representative value (e.g., the minimum or median). However, this method is not only costly but also unreliable. Extracting a single, meaningful value from a small set of noisy measurements is far from trivial, as the noise can follow irregular patterns. As a result, empirical models can reflect noise rather than the underlying behavior of the application.
An alternative strategy is to use noise-resilient priors, which guide the model fitting process by limiting it to certain function classes, such as polynomials or logarithms~\cite{calotoiu2013using, Goldsmith_ea:2007, Carrington_ea:2006}. These priors reflect the expectation rooted in algorithmic complexity theory. However, when these priors are applied statically, the degree of freedom they offer is still too large to effectively eliminate the influence of noise when run-to-run variation is high---especially if the performance model has multiple parameters~\cite{ritter2020learning}. 

To overcome these limitations, we propose a novel approach that introduces dynamic priors based on noise-resilient measurements of computational effort, combined with knowledge of typical algorithms used in communication routines. We call them dynamic because they are generated individually for each call path of an application based on effort metrics collected at runtime. They aim to further limit the search space of possible performance models to those that reflect this effort. Our approach relies on software counters that track the number of executed basic blocks, communication operations, and message sizes. Because basic blocks are free of branches, their execution cost is inherently constant, making their counts robust indicators of asymptotic computational complexity. 
We integrate our approach into \extrap{}, a state-of-the-art tool for empirical performance modeling. As illustrated in \cref{fig:method}, our method allows for more accurate and cost-efficient performance models better aligned with theoretical expectations. In summary, our key contributions are:

\begin{itemize} 

\item We reduce the influence of performance noise by integrating dynamic, application-specific priors derived from noise-resilient effort metrics. This significantly improves model accuracy while lowering the number of required experiments. 

\item  We design an open-source benchmark generator for synthetic parallel programs with predefined theoretical complexity,  enabling evaluation of our method with a sufficiently large population of benchmarks. 

\item We validate our approach using both synthetic benchmarks and realistic mini-applications. Under high-noise conditions, our method reduces the average prediction error from  $84\%$ to $20\%$, while cutting experimental costs in half. 
\end{itemize}

Before we explain our approach in more detail, we briefly introduce Extra-P, the tool we aim to improve, and its technical foundations.

\section{Background} \label{sec:background}

\extrap{}~\cite{calotoiu2013using} is a tool to automatically generate performance models for HPC applications using small-scale measurements. While its underlying method is programming-model agnostic, it has been demonstrated with MPI~\cite{calotoiu2013using}, OpenMP~\cite{iwainsky2015many}, CUDA~\cite{geiss_ea:2025}, and even with combinations of the three~\cite{geiss_ea:2025}. It considers $m$ input parameters (e.g., number of processors or input sizes), represented by $\{x_1, \ldots, x_m \}$. For model generation, we collect empirical data for combinations of input parameters using measurement tools like \scorep~\cite{scorep}.
Subsequently, \extrap{} derives human-readable performance equations from the measurements, exposing performance bottlenecks when the application is scaled.   
A core concept of \extrap{} is the \textit{performance model normal form} (PMNF)~\cite{calotoiu2016fast}. The PMNF for multiple parameters is defined by
\begin{equation} \label{eq:ep_multpar}
f(x_{1},\ldots,x_{m}) = \sum^{n}_{k=1} c_{k} \prod_{l=1}^{m} x_{l}^{i_{k_l}} \log_2^{j_{k_l}} (x_{l})
\end{equation}
where $n$ is the number of terms, $c_k$ are constant coefficients, and $i_{k_l}$ and $j_{k_l}$ are the exponents of the monomials and logarithms chosen from sets $I$ and $J$, respectively. 
The PMNF, together with the exponent sets $I$ and $J$, defines the search space of performance-model hypotheses, where the $n$ terms represent the influence of the parameters $\{x_1, \ldots, x_m \}$ as a combination of monomial and logarithmic expressions. By default, $n=m+1$ terms are used, reducing the risk of overfitting. One of these terms is always a constant. The PMNF serves as a prior that confines the space of discoverable models to functions that resemble complexity expressions typically found in algorithm textbooks. While already providing a certain degree of noise resilience, the space of model hypotheses remains still large enough to let noise adversely affect the model quality. Currently, the sets of possible exponents, $i_{k_l} \in I$ and $j_{k_l} \in J$, are manually chosen. The preset configuration, which we also use for this study, is shown below. %\RemarkFelix{I reverted the equation to the old version, but, now that Alexander mentions it, I am wondering why 4/5 is included while 1/5 is missing.} \RemarkGustavo{The term 1/5 is not included in the list of extra-p terms in the version we used.} \RemarkFelix{Is there a specific reason? At any rate, we should discuss the default after submission.}

\begin{subequations}
\label{eq:IandJ}
%\begin{equation}
%\label{eq:Ialt}
%\scalebox{0.942}{I = \{ i \mid i \in [0,3] \land i=\frac{n}{d} \land n \in \mathbb{N}_0 %\land d \in \{3,4,5\}\}},
%\end{equation}
\begin{equation}
\label{eq:I}
\scalebox{0.942}{$I =  \left\{  0, \frac{1}{4}, \frac{1}{3}, \frac{1}{2}, \frac{2}{3}, \frac{3}{4}, \frac{4}{5}, 1, \frac{5}{4}, \frac{4}{3}, \frac{3}{2}, \frac{5}{3}, \frac{7}{4}, 2, \frac{9}{4}, \frac{7}{3}, \frac{5}{2}, \frac{8}{3}, \frac{11}{4}, 3 \right\}$}
\end{equation}
\begin{equation}
\label{eq:J}
\scalebox{0.942}{$J =  \{ 0, 1, 2 \}$}
\end{equation}
\end{subequations}
The equation coefficients are determined using linear regression, and the optimal hypothesis is selected based on cross-validation. Thus, the measurements are the key component in finding the correct models. The more parameters are considered, the more freedom the modeling process has to fit the data, which increases the chances of errors caused by random noise or other interference such as network congestion~\cite{ritter2020learning}. 

Several heuristics have been devised to further reduce the impact of noise in \extrap{}. The best example is an adaptive modeling method, which achieves a certain degree of noise resilience by utilizing deep neural networks (DNNs)~\cite{ritter2021noise}. DNNs can be trained with noisy data to enhance noise resilience. This technique generates performance models either using the regression-based search or a DNN, depending on the function's characteristics and the estimated noise level. The DNN architecture includes an input layer with $11$ neurons, $5$ hidden layers ($2 \times 1500$, $750$, $2 \times 250$ neurons), with a tanh activation function. The output is a fully connected dense layer with $43$ neurons and a softmax activation function, which predicts the exponent combinations. Since the objectives of this DNN-based approach are identical to the ones we pursue in this work, we compare both experimentally in Section~\ref{sec:evaluation}. 

Another method is taint analysis, supported by LLVM, which identifies input parameters influencing performance~\cite{copik2021extracting}. This strategy aims to mitigate the influence of noise by excluding model functions from the search space with parameters that do not affect performance. \extrap{} has also taken advantage of noise-resilient metrics for specific applications of performance models. For example, performance models derived from task graphs helped identify parallel efficiency bounds based on the work-depth model~\cite{shudler2017isoefficiency}. Furthermore, hardware and software counters recording elementary operations related to different types of resources, such as the processor or the network, were used to create requirement models of applications aiding software-hardware co-design~\cite{calotoiu2018lightweight}. 
In addition, \extrap{} has been used to uncover deviations between the theoretical and actual scaling of MPI collective operations~\cite{shudler2019engineering}. The way we treat communication in this paper drew inspiration from this study. Similarly to this work, we leverage theoretical runtime bounds from the literature but scale them based on the number of bytes transferred.

\section{Approach} \label{sec:method}

Our approach introduces per-call-path dynamic priors that are inherently robust to noise and integrates them with conventional noise-sensitive measurements to build resilient performance models. This approach significantly reduces the adverse impact of noise during model generation, leading to models that better reflect the application’s theoretical behavior. As we will show later, this strategy also reduces the number of required experiments, lowering the overall cost of measurements.

Traditional performance measurement tools like \scorep{} often yield data with substantial run-to-run variability, making such measurements unreliable for constructing accurate models. To overcome this limitation, our workflow (illustrated in \cref{fig:method}) augments noise-sensitive data with noise-resilient effort measurements, collected via software counters using a modified version of \scorep{} (\cref{sec:method:measurements}). These measurements capture stable computation and communication effort indicators, such as the number of executed basic blocks and message sizes. From this data, we derive dynamic priors that reflect the structural cost of specific code (\cref{sec:method:priors}). Finally, we combine these resilient priors with the original, noise-sensitive timing data to guide the performance modeling process, resulting in final models that are both accurate and resilient to noise (\cref{sec:method:modeling}). To demonstrate the practical value of our approach, we implemented the modeling step within \extrap{}, showing how noise-resilient dynamic priors can enhance model quality under noise conditions.

\subsection{Noise-resilient measurements} \label{sec:method:measurements}

Measurement tools are essential for the performance analysis and optimization on HPC systems. One commonly used tool is \scorep~\cite{scorep}, a profiling and event-tracing tool. \scorep instruments code during compilation, dynamically measuring performance, tracing parallel call-path profiles, and reporting execution times. However, similar to many other conventional profiling tools, it measures wall-clock execution time to gauge performance, a metric vulnerable to noise.

To address this, we combine noise-sensitive metrics with noise-resilient effort metrics that remain stable across executions. For example, we count executed basic blocks, a unit of work that abstracts away timing fluctuations. A basic block is a straight-line code sequence with no internal branches, whose execution time (excluding function calls) is roughly constant. Counting them provides a reliable estimate of computational effort and supports asymptotic complexity analysis. Basic block executions can be easily counted using LLVM~\cite{LLVMwebsite}, enabling straightforward instrumentation and measuring code effort. 

We integrate a custom LLVM plug-in into the \scorep{} workflow to instrument code at the basic-block level. Unlike conventional time-based profiling, our tool counts basic-block executions per MPI rank for each function. While basic-block instrumentation introduces overhead---from our findings, the overhead in applied case studies can vary from around $60\%$ to $100\%$---it does not compromise accuracy. Unlike time-based profiling, our method is immune to such an intrusion, making it suitable for precise workload characterization and improved performance modeling. In addition to computational effort, we approximate communication effort by recording the number of processes involved and the bytes transferred during MPI operations. This data is derived from MPI-call arguments following MPI standards---for instance, an MPI\_Bcast of $n$ integers to $p$ processes records $p \cdot n \cdot 4$ bytes sent by the root and $n \cdot 4$ bytes received at each target. 
We use these stable measurements in the next section to construct priors for computational and communication effort, forming the foundation of our software--counter--based (SWC-based) performance modeling framework. 

\subsection{Deriving priors} \label{sec:method:priors}

The key idea of SWC-based performance modeling is to integrate additional noise-resilient priors into the modeling process. These priors are created from the SWC-based effort metrics, and they act as structural guides, capturing the underlying computational and communication effort based on noise-resilient effort measurements (i.e., basic-block counts and the number of transferred bytes). We start this process by using \extrap to generate performance models for each call path, reflecting how the effort scales with changing execution parameters. 

For the \textit{computational effort}, we expect that the asymptotic complexity derived from the basic-block model of a function accurately represents its scaling behavior. Consequently, we derive the prior by preserving the exponents of the basic-block model while stripping it of the coefficients. These will be added back in the next step to align the model with the execution time measurements. 

In contrast, for \textit{communication effort}, models based solely on transferred bytes are insufficient to fully capture performance, as the type of data transfer also impacts the outcome. Therefore, we integrate the model of transferred bytes, denoted as $B(p, \textbf{x})$, into well-established communication performance models for both point-to-point and collective operations (see~\cref{table:mpi_com}), as described in the literature~\cite{chan2007collective, zhang2017predicting}. These models incorporate key coefficients such as latency $\alpha$, bandwidth $\beta$, and computation cost $\gamma$ (defined using wall-clock measurements). As highlighted by the $\log_2(p)$ term in~\cref{table:mpi_com}, the models for collective communication account for the impact of the number of processes on transfer time, particularly due to step-by-step data transfers. This is based on the assumption that each MPI rank can only send one message at a time~\cite{chan2007collective}, leading to a tree-based transfer process for information sharing across ranks. Thus, the final prior approximates both the raw data transfer and the time necessary for the step-by-step transfer. %\RemarkChanges{Additionally, asynchronous MPI sends are represented analogously to synchronous sends, with the transferred bytes accounted for in the corresponding MPI\_Test calls.}

\begin{table}
\centering
\caption{Theoretical performance of MPI point-to-point and collective communication operations. $f(p, \textbf{x})$ is the expected runtime, $B(p, \textbf{x})$ are the bytes transferred, $p$ denotes the number of MPI ranks, $\textbf{x} = \{x_1, \ldots, x_m \}$ are the remaining performance-model parameters, and variables $\alpha$, $\beta$, and $\gamma$ correspond to latency including overhead, bandwidth, and computation cost.} 
%\begin{tabularx}{\linewidth}{lXl} 
\begin{tabular}{lll}
\toprule
\textbf{MPI function} & \textbf{Expected runtime} & \textbf{Ref.} \\  
\midrule
% Line 
Send      & $f(\textbf{x},p) = \alpha +  B(\textbf{x},p) \beta$ & \cite{zhang2017predicting} \\ 
% Line 
Receive   & $f(\textbf{x},p) = \alpha +  B(\textbf{x},p) \beta$ & \cite{zhang2017predicting} \\ 
% Line 
Broadcast     & $f(\textbf{x},p) = \log_2(p) \alpha + B(\textbf{x},p) \beta$ & \cite{chan2007collective} \\  
% Line 
Scatter   & $f(\textbf{x},p) = \log_2(p) \alpha + B(\textbf{x},p) \frac{p-1}{p} \beta$ & \cite{chan2007collective} \\ 
% Line 
Gather    & $f(\textbf{x},p) = \log_2(p) \alpha + B(\textbf{x},p) \frac{p-1}{p} \beta$  & \cite{chan2007collective} \\ 
% Line 
Allgather & $f(\textbf{x},p) = \log_2(p) \alpha + B(\textbf{x},p) \frac{p-1}{p} \beta$ & \cite{chan2007collective} \\ 
% Line 
Reduce    & $f(\textbf{x},p) = \log_2(p) \alpha + \left( \beta + \frac{p-1}{p} \gamma \right) B(\textbf{x},p) $  & \cite{chan2007collective} \\ 
% Line 
Allreduce & $f(\textbf{x},p) = \log_2(p) \alpha + \left( \beta + \frac{p-1}{p} \gamma \right) B(\textbf{x},p) $  & \cite{chan2007collective} \\ 
\bottomrule
%\end{tabularx}
\end{tabular}
\label{table:mpi_com}
\end{table}

\subsection{Creating models from priors} \label{sec:method:modeling}

In the final step, we combine the dynamic priors with noise-sensitive measurements to construct the SWC-based performance models. Specifically, each prior serves as a skeleton model---a functional form without fixed coefficients---which is then instantiated by fitting its coefficients to the noise-sensitive measurements using \extrap's regression-based modeler. The resulting model predicts runtime while preserving the scaling behavior dictated by the underlying noise-resilient effort. This approach ensures robustness against noise while maintaining an intuitive link to actual runtime behavior.

Because these SWC-based models reflect only the active computational and communication effort, they do not capture waiting time. The only exception is busy waiting, where a process repeatedly executes a loop (i.e., spin-waits) until a condition is met. In such cases, each loop iteration contributes to the basic-block count. As a result, when the basic-block metric is used to construct the computational prior, this prior will also account for the waiting time caused by busy waiting. This can lead to priors and, consequently, performance models that are partially influenced by system noise.

\subsection{Example} 

\begin{figure} 
\centerline{\includegraphics[width=\linewidth]{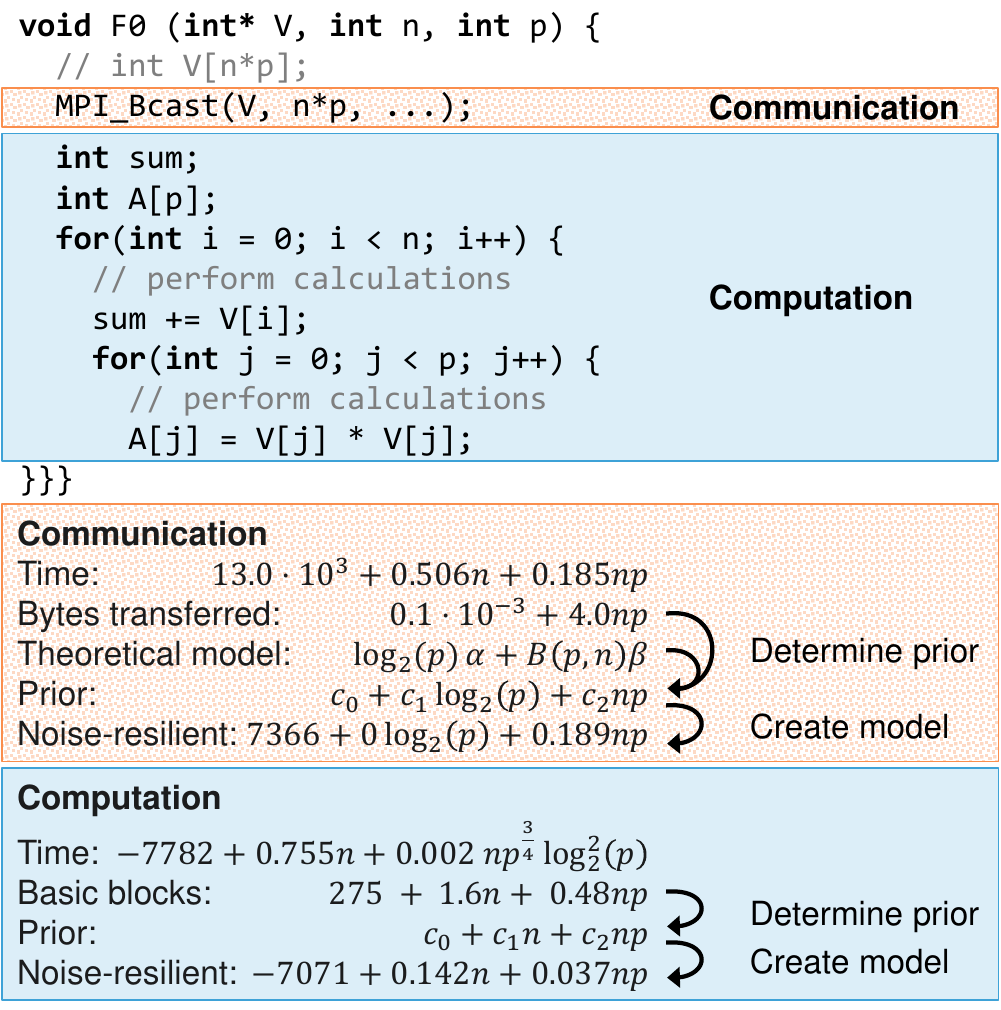}}
\caption{Our SWC-based performance modeling approach applied to an example function. Besides the noise-sensitive time model and the final SWC-based model, we show the intermediate models and priors for communication and computation.} 
\label{fig:example}
\end{figure}

\Cref{fig:example} illustrates the complete modeling workflow---including intermediate models and priors---using a synthetic example. We begin by executing a code snippet while varying both the input size $n$ and the number of processes $p$, collecting measurements for runtime, basic-block counts, and transferred bytes. We first construct intermediary models for basic-block counts and bytes transferred from these measurements. These are then used to derive the corresponding priors. Finally, we determine the coefficients of the priors by fitting them to the time measurements, resulting in a SWC-based performance model. We use \extrap to create these performance models. Data collection and performance modeling are performed individually for each call path, enabling us to pinpoint scaling bottlenecks in the source code for noise-resilient and noise-sensitive measurements.

In this example, the function F0 contains a nested loop and includes a single call to MPI\_Bcast. The basic-block model successfully captures the asymptotic complexity of the computation inside the function, $\mathcal{O}(n+np)$, and this structural behavior is directly reflected in the derived prior. In contrast, the time-based model is affected by noise, causing deviations in the captured complexity. This is especially apparent in multi-parameter scenarios, where the influence of one parameter (e.g., input size) may obscure the effect of another (e.g., number of processes). Our method makes such inconsistencies visible by comparing the time-based model to the prior, allowing us to correct them and produce a time model that better matches the true asymptotic scaling. We follow a similar process for the communication function: we first model the transferred bytes and then combine this model with the theoretical performance model for MPI\_Bcast (see~\cref{table:mpi_com}) to derive the prior. Again, we observe that the time-based model deviates from the expected complexity, while the prior maintains alignment with theoretical behavior.

Beyond mitigating the effects of measurement noise, our approach also reduces the overall experimental cost. Since the prior stabilizes the model structure, much fewer time measurements are needed to fit the final model accurately. The overhead of collecting one additional noise-resilient metric per parameter configuration is thus easily offset. 

\section{Benchmark Generator} \label{sec:bg}

Numerous benchmarks for HPC applications have been proposed over time~\cite{cascajo2024detecting, yan2023synthesizing, chamzas2021motionbenchmaker}. 
However, their analytical complexity is usually either unknown or undocumented, and benchmarks with known complexity are so rare that they do not cover all possible complexity classes. 
To overcome these limitations, we introduce a benchmark generator for parallel codes that can capture a broader range of algorithmic complexities. Users can specify both computational and communication complexities, and the generator automatically reproduces them by implementing loop constructs and message sizes with controlled scaling behavior, along with diverse computation operations and MPI functions. For this, we group multiple C++ kernels into a benchmark-frame function that sets up the specific environment by allocating memory and executing the code. The user can define the kernels' complexities, or they can be chosen randomly. We use combinations of linear and nonlinear terms to represent complexities:
\begin{equation} \label{eq:nonli}
F = \left\{ x^{i}, \log_2^{j}(x)\right\}, 
\end{equation}
where~$x$ represents an execution parameter of the benchmark run (e.g., the problem size~$n$ or the number of MPI ranks~$p$), and $i \in I$ and $j \in J$ as in~\cref{eq:IandJ}. The exponents presented in~\cref{eq:IandJ} reflect complexities typically found in HPC applications, while those that do not belong to this set are unlikely to be found in practice. The user has the flexibility to extend the sets of functions and exponents. If the complexities are generated randomly, they are composed using random combinations of terms from~$F$. 

For the communication aspect of the kernel, the benchmark generator utilizes various MPI library functions, including allgather, allreduce, barrier, broadcast, gather, reduce, scatter, send, and receive. The expected performance of these functions is determined by the frequency of their execution and the size of the transmitted messages, both of which are controlled by elements of~$F$, as shown in~\cref{eq:nonli}. 
A key aspect of our benchmark is using arrays whose sizes are derived from the overall problem size by applying a user-defined scaling behavior. This allows users to specify how array sizes grow or shrink with problem size. The resulting array size directly relates to the amount of data that must be transferred, thereby influencing communication cost. By combining this with the selection of MPI library functions, users have fine-grained control over the communication complexity of the benchmark. 
As these arrays are transmitted over the network, users can also control communication complexity.
We follow a similar approach for computational performance: The benchmark generator inserts loops with ranges determined by elements of~$F$ and arranges these loops in either sequential or nested configurations. Subsequently, the arrays used in the communication step are subjected to operations, such as addition or multiplication, within the loops comprising the computation. After generating the code, we instrument it using our customized version of \scorep. 

\begin{figure*} 
\centerline{\includegraphics[width=\linewidth]{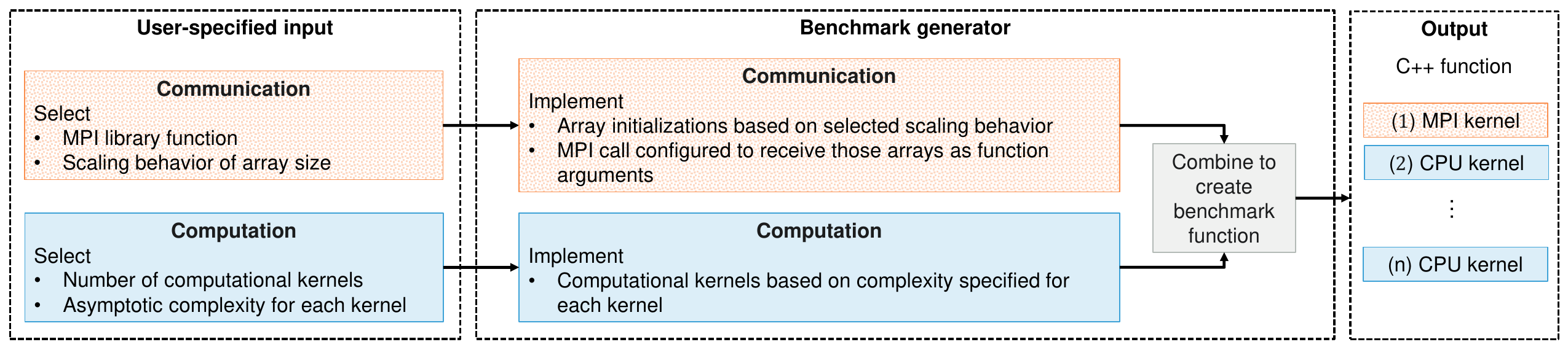}}
\caption{The user selects an MPI function and specifies the scaling behavior of the message volume and the computation phases following the communication. In the following step, the generator reproduces this behavior by combining defined code templates that capture the intended performance characteristics. Finally, the generator composes C++ code employing these characteristics.} 
\label{fig:bgwork}
\end{figure*}

\Cref{fig:bgwork} shows the benchmark generator workflow. The user selects a communication scaling behavior along with an MPI function, in addition to computational kernels with different scaling behaviors, and whether they should be arranged in a nested or sequential manner in the code. The computational complexity is achieved by the loop arrangements, and the communication complexity by the MPI function. The generator then composes benchmarks exhibiting the intended scaling characteristics. As an example, \cref{fig:example} illustrates a benchmark function with two execution parameters: the number of MPI ranks $p$ and the input size $n$. The communication performance is driven by an MPI\_Bcast operation that distributes a vector containing $n \cdot p$ integer values. Computation is performed within two nested loops; the outer one, iterating from $0$ to $n$, does vector reduction and contains the inner loop, iterating from $0$ to $p$, performing vector multiplication. The code exhibits a communication complexity of $\mathcal{O}(n\cdot p + \log_2 (p))$ and a computational complexity of $\mathcal{O}(n\cdot p+n)$. This example demonstrates how the function's complexity is managed. The collective operation determines the expected communication complexity, as outlined in \cref{table:mpi_com}, while the nested loops control the computational complexity.

Below, we use the benchmark generator to validate empirical performance models derived from measured performance. After profiling and modeling runtime performance, the generated benchmarks help identify deviations between theoretical expectations and the generated empirical performance models, allowing the accuracy of the models to be gauged.

\section{Evaluation} \label{sec:evaluation}

After introducing the evaluation criteria and methodology, we present our results---first, those obtained with synthetic benchmarks, followed by case studies with realistic applications.

\subsection{Evaluation methodology} \label{sec:eval:method}

Our evaluation methodology comprises seven elements: (1)~evaluation criteria, (2)~comparison baseline, (3)~synthetic benchmark generator, (4)~test systems, (5)~measurement selection, (6)~model accuracy metrics, and (7)~artificial noise characteristics. 

\subsubsection{Criteria} \label{sec:eval:method:criteria}

We evaluate our approach using the following three criteria:
\textbf{(i)~Accuracy.} How close are the resulting models to theoretical expectations, and how well can they predict actual runtime?
\textbf{(ii)~Robustness to noise.} How well can our models preserve accuracy in the presence of varying degrees of noise? 
\textbf{(iii)~Experimental cost.} How much cheaper is our new approach regarding the number of times performance experiments have to be repeated?

\subsubsection{Comparison baseline} \label{sec:eval:method:baseline}

To assess the effectiveness of our \textbf{SWC-based} approach, which stabilizes noise-sensitive time measurements using noise-resilient priors, as discussed in Section~\ref{sec:method}, we compare it with the two alternatives mentioned in Section~\ref {sec:background}: \textbf{classic} modeling using only execution-time measurements and the \textbf{DNN-based} method, which uses machine learning to mitigate the noise effect~\cite{ritter2021noise}. 

We use \textbf{theoretical} models derived from analytical performance to establish a reliable reference. These models serve as the ground truth for assessing the accuracy and fidelity of each performance modeling approach. For this purpose, we employ a synthetic benchmark generator to extend the evaluation across a broader spectrum of pre-defined algorithmic complexities (as outlined below). Furthermore, we utilize realistic HPC applications whose theoretical models have been presented in prior studies.

\subsubsection{Test systems} \label{sec:eval:method:config}

The experiments are run on two systems. The first is the MPI section of the cluster Lichtenberg II Phase $1$ at TU Darmstadt.
Each node of the section runs on Red Hat Enterprise Linux 8.8 and consists of two Intel Xeon Platinum 9242 CPUs with 48 cores at 2.3~GHz each and 384~GB of RAM. The nodes are connected with Infiniband HDR100 in a 1:1 non-blocking configuration. 
The second system is the CPU section of the cluster Jureca-DC at the J\"ulich Supercomputing Centre. 
Each node has two AMD EPYC 7742 processors with 64 cores each and 512~GB of DDR4 RAM. The nodes are connected via Infiniband HDR100.

\subsubsection{Selection of measurements} \label{sec:eval:method:measurements}

For each model, we select $5$ representative values per parameter. This leads to $5^{m}$ independent execution configurations for a model with $m$ parameters. In the classic (noise-sensitive) approach, each configuration is measured $5$ times to mitigate noise, and the median is considered as the input value. This results in a total of $5^{m+1}$ measurements. Conversely, our proposed approach combines a single noise-sensitive measurement per configuration with additional noise-resilient metrics, reducing the total measurement count to $2 \cdot 5^{m}$. 

Of course, considering all combinations of parameter values is computationally expensive. Therefore, a method called sparse modeling~\cite{ritter2020learning} has been developed to reduce experimental costs by requiring fewer data points. However, the exact number of measurements needed for sparse modeling is not systematically defined, which makes it hard to compare the noise-resilient and the classic methods. In future work, we plan to extend our analysis to incorporate sparse modeling and explore the optimal number of data points required for its effectiveness when following the noise-resilient strategy.

\subsubsection{Accuracy metrics} \label{sec:eval:method:metrics}

We assess model accuracy using two metrics. The first, the \textit{exponent deviation}~(ED), aligns with the soft-O ($\mathcal{\tilde{O}}$) notation~\cite{cormen2022introduction} and assesses the similarity between performance models in terms of their asymptotic complexity:
\begin{equation} \label{eq:deviation}
\text{ED}(f_1(x_{i}), f_2(x_{i}))  = \left| n_1 - n_2 \right|,
\end{equation}
For logarithmic or constant terms, an exponent of $0$ is assigned (e.g., $\log_2(x)$ is equivalent to $x^{0}\log_2(x)$). This metric indicates how closely the asymptotic complexities of different models align. Small exponent deviation values are desirable, with an exponent deviation equal to $0$ signifying that the growth rates of the evaluated functions are in close agreement.

The second metric, the \textit{relative error}~(RE), quantifies the predictive power of a performance model concerning an unseen test point: 
\begin{equation} \label{eq:tpe}
\text{RE}(f_1(x_{i})) = \frac{|y_{i} - f_1(x_{i})|}{y_{i}} \cdot 100 \%, 
\end{equation}
where $f_1(x_{i})$ represents the model's predicted value, while $y_{i}$ denotes the measured value at the test point, both corresponding to the input $x_{i}$.  
In our analysis, the test point is the next data point beyond the range used for training, following the same interval rule. For instance, if the training points consist of $\{2^{5}$, $2^{6}$, $2^{7}$, $2^{8}$, $2^{9}\}$, the test point is $2^{10}$. Smaller relative error values are desirable, with $0\%$ indicating an ideal model. 

\subsubsection{Artificial noise} \label{sec:eval:method:an}

We introduced artificial noise into randomly selected runtime measurements as $\tilde{y} = y + \eta$, where $y$ is the original measurement and $\eta$ the sampled noise. The artificial noise $\eta$ has varying intensities and is chosen from five different noise patterns: uniform distribution ($a = 0$, $b = 1$), truncated normal distribution ($\mu = 0$, $\sigma = 1$, $a = 0$, $b = 1$), scaled Poisson distribution ($\lambda=1000$, scale parameter $=1000$), scaled exponential distribution ($\lambda=1000$, scale parameter $=1000$), and no noise. The noise distribution is clipped to $[0, 1]$. For the sake of simplicity, we restrict ourselves to additive noise ($\eta \ge 0$) because the impact of subtractive noise is negligible. The resulting data is designed to emulate real-world random perturbations and is used to evaluate the robustness of our method.

% ----------------------------------------------------------
% Synthetic evaluation
% ----------------------------------------------------------

\subsection{Evaluation with synthetic benchmarks} \label{sec:eval:synthetic}

For this evaluation, we generated synthetic benchmarks with a total of $200$ functions, which we executed on  Lichtenberg II, meaning that only the benchmark codes are synthetic, while we collect real measurements of runtime, basic-block counts, and transferred bytes. 
Our evaluation focuses on weak-style scaling performance models involving two primary variables: the number of MPI ranks~($p$) and the problem size~($n$). The training parameters are defined with $p \in \{128$, $256$, $512$, $1024$, $2048\}$, and $n \in \{ 8000$, $16000$, $24000$, $32000$, $40000\}$. The test point is $p = 4096$ and $n = 48000$. 

The DNN-based modeler failed for our synthetic benchmarks almost completely, deriving constant models across the board where non-constant relationships exist. The reason is that it has not been designed for functions with such short runtimes. As the input parameters $n$ and $p$ vary, our benchmarks exhibit only minor runtime variations. In a perfect, noise-free system, these input changes would produce clearly distinguishable runtime differences. However, even the natural background noise on our test system masks these variations. As a result, the DNN-based modeler perceives the input-output differences as insignificant, leading to constant models. In contrast, classic modeling better captures such subtle variations, whereas the input-output variations become even more evident in noise-resilient measurements. Below, we therefore restrict the detailed quantitative analysis of our synthetic benchmarks to classic and SWC-based models, because only those could identify meaningful relationships between model parameters and performance. Nevertheless, we provide a quantitative comparison of all three when we discuss realistic benchmarks in Section~\ref{sec:eval:real}.

\subsubsection{Accuracy} \label{sec:eval:synthetic:accuracy}

We first compare our SWC-based models with classic models by analyzing deviations in ED against ground-truth complexity (\cref{eq:deviation}). Without artificial noise, SWC-based models perfectly match theoretical growth (ED $= 0$ in $100\%$ of cases), while classic models often vary, particularly for small functions where system noise dominates. \Cref{table:rq1} summarizes the average ED, showing our models maintain zero deviation, in contrast to substantial errors in classic models.

We further assess the predictive power at the test points using the RE (\cref{eq:tpe}). Classic models yield average errors of $45\%$ (computation) and $91\%$ (communication), whereas SWC-based models achieve $35\%$ and $60\%$, respectively---a clear improvement. The communication error is higher due to variability from synchronization delays, which are not captured by effort metrics. Note that the test point covers twice the range of the training setup in terms of MPI ranks. Consequently, the results show relatively high errors in communication. Nevertheless, SWC-based communication models still accurately capture the order of magnitude, which is sufficient to identify bottlenecks.

\begin{table}
\centering
\caption{Average exponent deviations when comparing SWC-based and classic models against their theoretical expectations.}
\resizebox{\columnwidth}{!}{\begin{tabular}{l l l l}
\toprule
\textbf{Models} & \textbf{Computation} & \multicolumn{2}{c}{\textbf{Communication}} \\
&  & \textbf{MPI ranks} & \textbf{Message size}  \\
\midrule
SWC-based & $0$ & $0$ & $0$ \\
Classic & $0.44$ & $1.14$ & $0.57$  \\
\bottomrule
\end{tabular}}
\label{table:rq1}
\end{table}

\subsubsection{Robustness to noise} \label{sec:eval:synthetic:robust}

To evaluate robustness, we inject artificial noise into runtime data at intensities of $2\%$, $5\%$, $10\%$, $50\%$, and $75\%$. The randomly generated noise is added to the training data, which is then used to create new performance models. We apply $100$ independent random noise values for each function and noise pattern. Our SWC-based approach leaves the exponents fixed, keeping an ED $= 0$ across all noise levels. In contrast, exponent deviation increases substantially in classic models (\cref{fig:noise_bg_comp_ED,fig:noise_bg_mpi_ED}). We also evaluate RE under noise (\cref{fig:noise_bg_comp_RE,fig:noise_bg_mpi_RE}). While classic models' errors peak at $55\%$ (computation) and $127\%$ (communication), our method restricts maximum errors to $35\%$ and $60\%$, demonstrating strong robustness. In the figures, the function's asymptotic complexity, based on noise-resilient priors, remains stable despite noise (ED $= 0$). On the other hand, noise still affects our models when validating the results using RE values. Because runtime measurements are used to define the coefficients of the equations, noise can affect the predictive power. However, the RE values are significantly lower than those of classic methods.

\begin{figure}
\centering
  \begin{subfigure}[t]{0.48\linewidth}
    \centering
    \resizebox{\columnwidth}{!}{%
        \makeatletter
\newcommand\HUGE{\@setfontsize\Huge{25}{25}}
\newcommand\MID{\@setfontsize\Mid{20}{20}}
\makeatother

\begin{tikzpicture}
  \centering
  \begin{axis}[
        height=8.5cm, width=11cm,
        bar width=1.0cm,
        ymajorgrids, tick align=inside,
        major grid style={draw=gray!20!white},
        enlarge y limits={value=.1,upper},
        ymin=0, ymax=1.0,
        axis x line*=bottom,
        axis y line*=left,
        y axis line style={opacity=0},
        tickwidth=0pt,
        enlarge x limits=true,
        legend style={
            draw=none,
            at={(0.5,-0.3)},
            anchor=center,
            legend columns=-1,
            mark size=6,
            mark options={scale=0.6},
            /tikz/every even column/.append style={column sep=0.5cm}            
        },
        legend to name={legend_noise_bg},
        tick label style = {font=\MID}, 
        ylabel={Exponent deviation},
        ylabel style={at={(-0.06,0.5)}},
        ylabel style={font=\HUGE},
        xlabel={Artificial noise},
        xlabel style={at={(0.5,-0.03)}},
        xlabel style={font=\HUGE},
        symbolic x coords={
           2\%, 5\%, 10\%, 50\%, 75\%},
       xtick=data,
       nodes near coords={}
    ]       
    
    % --------------------------------------------------------
    % Noise-sensitive
    \addplot [
        only marks, 
        mark=square*, 
        mark size=6,
        color=orange!70!white, 
        ] 
      coordinates {
      (2\%, 0.443990) 
      (5\%, 0.442393) 
      (10\%, 0.439011) 
      (50\%,  0.444195) 
      (75\%, 0.450842)};
      \addlegendentry[color=black]{Classic}        
    % --------------------------------------------------------

    % --------------------------------------------------------
    % Noise-resilient
    \addplot [
        only marks, 
        mark=*, 
        mark size=6,
        blue!70!white
        ] 
      coordinates {
      (2\%, 0.0) 
      (5\%, 0.0) 
      (10\%, 0.0) 
      (50\%, 0.0) 
      (75\%, 0.0)};
      \addlegendentry[color=black]{SWC-based} 
    % --------------------------------------------------------

    % --------------------------------------------------------
    % Noise-resilient
    \addplot [
        color=blue!70!white,          
        dashed,             
        line width=1pt     
        ]   
      coordinates {
      (2\%, 0.0) 
      (5\%, 0.0) 
      (10\%, 0.0) 
      (50\%, 0.0) 
      (75\%, 0.0)};

    % Noise-sensitive
    \addplot [
        color=orange!70!white,        
        dashed,             
        line width=1pt     
        ]   
      coordinates {
      (2\%, 0.443990) 
      (5\%, 0.442393) 
      (10\%, 0.439011) 
      (50\%,  0.444195) 
      (75\%, 0.450842)};
   
  \end{axis}
\end{tikzpicture}
    }
    \caption{Benchmark--computation}\label{fig:noise_bg_comp_ED}
  \end{subfigure}  
\hfill
 \begin{subfigure}[t]{0.48\linewidth} 
    \centering
    \resizebox{\columnwidth}{!}{%
        \makeatletter
\newcommand\HUGE{\@setfontsize\Huge{25}{25}}
\newcommand\MID{\@setfontsize\Mid{20}{20}}
\makeatother

\begin{tikzpicture}
  \centering
  \begin{axis}[
        height=8.5cm, width=11cm,
        bar width=1.0cm,
        ymajorgrids, tick align=inside,
        major grid style={draw=gray!20!white},
        enlarge y limits={value=.1,upper},
        ymin=0, ymax=140,
        axis x line*=bottom,
        axis y line*=left,
        y axis line style={opacity=0},
        tickwidth=0pt,
        enlarge x limits=true,
        legend style={
            draw=none,
            at={(0.5,-0.3)},
            anchor=center,
            legend columns=-1,
            /tikz/every even column/.append style={column sep=0.5cm}            
        },
        %legend to name={legend_noise},
        tick label style = {font=\MID}, 
        ylabel={Relative error (\%)},
        ylabel style={at={(-0.06,0.5)}},
        ylabel style={font=\HUGE},
        xlabel={Artificial noise},
        xlabel style={at={(0.5,-0.03)}},
        xlabel style={font=\HUGE},
        symbolic x coords={
           2\%, 5\%, 10\%, 50\%, 75\%},
       xtick=data,
       nodes near coords={}
       %nodes near coords={
       % \pgfmathprintnumber[precision=0]{\pgfplotspointmeta}
       %}
    ]
    
   % --------------------------------------------------------
    % Noise-sensitive
    \addplot [
        only marks, 
        mark=square*, 
        mark size=6,
        color=orange!70!white,
        ] 
      coordinates {
      (2\%, 46.518310) 
      (5\%, 47.421675) 
      (10\%, 48.161871) 
      (50\%, 51.329171) 
      (75\%, 55.611829)};
    % --------------------------------------------------------

    % --------------------------------------------------------
    % Noise-resilient
    \addplot [
        only marks, 
        mark=*, 
        mark size=6,
        blue!70!white
        ] 
      coordinates {
      (2\%, 35.183984) 
      (5\%, 35.031751) 
      (10\%, 34.822101) 
      (50\%, 34.271993) 
      (75\%, 34.369034) };
    % --------------------------------------------------------

    % --------------------------------------------------------
    % Noise-resilient
    \addplot [
        color=blue!70!white,          
        dashed,             
        line width=1pt     
        ]   
      coordinates {
      (2\%, 35.183984) 
      (5\%, 35.031751) 
      (10\%, 34.822101) 
      (50\%, 34.271993) 
      (75\%, 34.369034) };
    % --------------------------------------------------------

    % --------------------------------------------------------
    % Noise-sensitive
    \addplot [
        color=orange!70!white,     
        dashed,             
        line width=1pt     
        ]   
      coordinates {
      (2\%, 46.518310) 
      (5\%, 47.421675) 
      (10\%, 48.161871) 
      (50\%, 51.329171) 
      (75\%, 55.611829)};
    % --------------------------------------------------------
    
  \end{axis}
\end{tikzpicture}

				
    }
    \caption{Benchmark--computation}\label{fig:noise_bg_comp_RE}
  \end{subfigure} 
 \medskip 
\vspace{1ex}
 \begin{subfigure}[t]{0.48\linewidth} 
    \centering
    \resizebox{\columnwidth}{!}{%
        \makeatletter
\newcommand\HUGE{\@setfontsize\Huge{25}{25}}
\newcommand\MID{\@setfontsize\Mid{20}{20}}
\makeatother

\begin{tikzpicture}
  \centering
  \begin{axis}[
        height=8.5cm, width=11cm,
        bar width=1.0cm,
        ymajorgrids, tick align=inside,
        major grid style={draw=gray!20!white},
        enlarge y limits={value=.1,upper},
        ymin=0, ymax=1.0,
        axis x line*=bottom,
        axis y line*=left,
        y axis line style={opacity=0},
        tickwidth=0pt,
        enlarge x limits=true,
        legend style={
            draw=none,
            at={(0.5,-0.3)},
            anchor=center,
            legend columns=2,
            /tikz/every even column/.append style={column sep=0.5cm}            
        },
        tick label style = {font=\MID}, 
        ylabel={Exponent deviation},
        ylabel style={at={(-0.06,0.5)}},
        ylabel style={font=\HUGE},
        xlabel={Artificial noise},
        xlabel style={at={(0.5,-0.03)}},
        xlabel style={font=\HUGE},
        symbolic x coords={
           2\%, 5\%, 10\%, 50\%, 75\%},
       xtick=data,
       nodes near coords={}
    ]       
    
    % --------------------------------------------------------
    % Noise-sensitive
    \addplot [
        only marks, 
        mark=square*, 
        mark size=6,
        color=orange!70!white,
        ] 
      coordinates {
      (2\%, 0.835643) 
      (5\%, 0.838447) 
      (10\%, 0.842607) 
      (50\%, 0.857864) 
      (75\%, 0.871264)};     
    % --------------------------------------------------------

    % --------------------------------------------------------
    % Noise-resilient
    \addplot [
        only marks, 
        mark=*, 
        mark size=6,
        blue!70!white
        ] 
      coordinates {
      (2\%, 0.0) 
      (5\%, 0.0) 
      (10\%, 0.0) 
      (50\%, 0.0) 
      (75\%, 0.0)};
    % --------------------------------------------------------

    % --------------------------------------------------------
    % Noise-resilient
    \addplot [
        color=blue!70!white,          
        dashed,             
        line width=1pt     
        ]   
      coordinates {
      (2\%, 0.0) 
      (5\%, 0.0) 
      (10\%, 0.0) 
      (50\%, 0.0) 
      (75\%, 0.0)};
    % --------------------------------------------------------

    % --------------------------------------------------------
    % Noise-sensitive
    \addplot [
        color=orange!70!white,          
        dashed,             
        line width=1pt     
        ]   
      coordinates {
      (2\%, 0.835643) 
      (5\%, 0.838447) 
      (10\%, 0.842607) 
      (50\%, 0.857864) 
      (75\%, 0.871264)};
    % --------------------------------------------------------
   
  \end{axis}
\end{tikzpicture}
    }
    \caption{Benchmark--communication}\label{fig:noise_bg_mpi_ED}
  \end{subfigure}
\hfill
 \begin{subfigure}[t]{0.48\linewidth} 
    \centering
    \resizebox{\columnwidth}{!}{%
        \makeatletter
\newcommand\HUGE{\@setfontsize\Huge{25}{25}}
\newcommand\MID{\@setfontsize\Mid{20}{20}}
\makeatother

\begin{tikzpicture}
  \centering
  \begin{axis}[
        height=8.5cm, width=11cm,
        bar width=1.0cm,
        ymajorgrids, tick align=inside,
        major grid style={draw=gray!20!white},
        enlarge y limits={value=.1,upper},
        ymin=0, ymax=140,
        axis x line*=bottom,
        axis y line*=left,
        y axis line style={opacity=0},
        tickwidth=0pt,
        enlarge x limits=true,
        legend style={
            draw=none,
            at={(0.5,-0.3)},
            anchor=center,
            legend columns=-1,
            /tikz/every even column/.append style={column sep=0.5cm}            
        },
        %legend to name={legend_noise},
        tick label style = {font=\MID}, 
        ylabel={Relative error (\%)},
        ylabel style={at={(-0.06,0.5)}},
        ylabel style={font=\HUGE},
        xlabel={Artificial noise},
        xlabel style={at={(0.5,-0.03)}},
        xlabel style={font=\HUGE},
        symbolic x coords={
           2\%, 5\%, 10\%, 50\%, 75\%},
       xtick=data,
       nodes near coords={}
    ]
    
   % --------------------------------------------------------
    % Noise-sensitive
    \addplot [
        only marks, 
        mark=square*, 
        mark size=6,
        color=orange!70!white, 
        ] 
      coordinates {
      (2\%, 93.128019) 
      (5\%, 95.307450) 
      (10\%, 98.283937) 
      (50\%, 112.527497) 
      (75\%, 127.402591)};
    % --------------------------------------------------------

    % --------------------------------------------------------
    % Noise-resilient
    \addplot [
        only marks, 
        mark=*, 
        mark size=6,
        blue!70!white
        ] 
      coordinates {
      (2\%, 60.676) 
      (5\%, 60.676) 
      (10\%, 60.676) 
      (50\%, 60.676) 
      (75\%, 60.676) };
    % --------------------------------------------------------

    % --------------------------------------------------------
    % Noise-resilient
    \addplot [
        color=blue!70!white,          
        dashed,             
        line width=1pt     
        ]   
      coordinates {
      (2\%, 60.676) 
      (5\%, 60.676) 
      (10\%, 60.676) 
      (50\%, 60.676) 
      (75\%, 60.676) };
    % --------------------------------------------------------

    % --------------------------------------------------------
    % Noise-sensitive
    \addplot [
        color=orange!70!white,     
        dashed,             
        line width=1pt     
        ]   
      coordinates {
      (2\%, 93.128019) 
      (5\%, 95.307450) 
      (10\%, 98.283937) 
      (50\%, 112.527497) 
      (75\%, 127.402591)};
    % --------------------------------------------------------
    
  \end{axis}
\end{tikzpicture}

				
    }
    \caption{Benchmark--communication}\label{fig:noise_bg_mpi_RE}
  \end{subfigure} 
  \ref{legend_noise_bg}

  \caption{Noise resilience of our SWC-based approach compared to standard classic modeling (i.e., without noise-resilient priors) for the generated benchmarks. Artificial noise of increasing intensity was added to the time measurements to simulate noisy conditions. The figure displays the average values of our evaluation metrics, the exponent deviation and the relative error between the predicted and measured values at the test points, computed over 100 independent noise-injection experiments.}  
  \label{fig:bg_noise}
\end{figure}

\subsubsection{Experimental costs} \label{sec:eval:synthetic:cost}

The simplicity of the benchmark functions ensures minimal instrumentation overhead. To assess the overhead of our noise-resilient basic-block counters in \scorep (\cref{sec:method:measurements}), we compare the execution time of the benchmark codes using either the standard and our modified \scorep version. Our results indicate that, on average, there is no significant difference in execution time between the two versions. 

Classical modeling requires five repetitions to reduce noise, and we analyze the impact of measurement repetition on model stability. We generate models using all combinations of $1$–$5$ measurement repetitions and evaluate deviations using ED (\cref{fig:reduction_bg_comp_ED,fig:reduction_bg_mpi_ED}) and RE (\cref{fig:reduction_bg_comp_RE,fig:reduction_bg_mpi_RE}). SWC-based models maintain an ED $= 0$ regardless of repetition count, whereas classic models show increasing deviations. The RE is also more stable in our models. Thus, our method achieves reliable performance models with only one repetition, reducing required measurements to $50$ ($25$ runtime, $25$ effort) compared to $125$ in traditional setups---less than half the experimental cost. 

\begin{figure}
\centering
  \begin{subfigure}[t]{0.48\linewidth}
    \centering
    \resizebox{\columnwidth}{!}{%
        \makeatletter
\newcommand\HUGE{\@setfontsize\Huge{25}{25}}
\newcommand\MID{\@setfontsize\Mid{20}{20}}
\makeatother

\begin{tikzpicture}
  \centering
  \begin{axis}[
        height=8.5cm, width=11cm,
        bar width=1.0cm,
        ymajorgrids, tick align=inside,
        major grid style={draw=gray!20!white},
        enlarge y limits={value=.1,upper},
        ymin=0, ymax=1.2,
        axis x line*=bottom,
        axis y line*=left,
        y axis line style={opacity=0},
        tickwidth=0pt,
        enlarge x limits=true,
        legend style={
            draw=none,
            at={(0.5,-0.3)},
            anchor=center,
            legend columns=-1,
            mark size=6,
            mark options={scale=0.6},
            /tikz/every even column/.append style={column sep=0.5cm}            
        },
        legend to name={legend_reduction_bg},
        tick label style = {font=\MID}, 
        ylabel={Exponent deviation},
        ylabel style={at={(-0.06,0.5)}},
        ylabel style={font=\HUGE},
        xlabel={Number of repetitions},
        xlabel style={at={(0.5,-0.03)}},
        xlabel style={font=\HUGE},
        symbolic x coords={
           1, 2, 3, 4, 5},
       xtick=data,
       nodes near coords={}
    ]       
    
    % --------------------------------------------------------
    % Noise-sensitive
    \addplot [
        only marks, 
        mark=square*, 
        mark size=6,
        color=orange!70!white,
        ] 
      coordinates {
      (1, 0.34)
      (2, 0.24)
      (3, 0.29)
      (4, 0.24)
      (5, 0.44)};
      \addlegendentry[color=black]{Classic}        
    % --------------------------------------------------------

    % --------------------------------------------------------
    % Noise-resilient
    \addplot [
        only marks, 
        mark=*, 
        mark size=6,
        blue!70!white
        ] 
      coordinates {
      (1, 0)
      (2, 0)
      (3, 0)
      (4, 0)
      (5, 0)};
      \addlegendentry[color=black]{SWC-based} 
    % --------------------------------------------------------

    % --------------------------------------------------------
    % Noise-resilient
    \addplot [
        color=blue!70!white,          
        dashed,             
        line width=1pt     
        ]   
      coordinates {
      (1, 0)
      (2, 0)
      (3, 0)
      (4, 0)
      (5, 0)};

    % Noise-sensitive
    \addplot [
        color=orange!70!white,        
        dashed,             
        line width=1pt     
        ]   
      coordinates {
      (1, 0.34)
      (2, 0.24)
      (3, 0.29)
      (4, 0.24)
      (5, 0.44)};
   
  \end{axis}
\end{tikzpicture}
    }
    \caption{Benchmark--computation}\label{fig:reduction_bg_comp_ED}
  \end{subfigure}  
\hfill
 \begin{subfigure}[t]{0.48\linewidth} 
    \centering
    \resizebox{\columnwidth}{!}{%
        \makeatletter
\newcommand\HUGE{\@setfontsize\Huge{25}{25}}
\newcommand\MID{\@setfontsize\Mid{20}{20}}
\makeatother

\begin{tikzpicture}
  \centering
  \begin{axis}[
        height=8.5cm, width=11cm,
        bar width=1.0cm,
        ymajorgrids, tick align=inside,
        major grid style={draw=gray!20!white},
        enlarge y limits={value=.1,upper},
        ymin=0, ymax=120,
        axis x line*=bottom,
        axis y line*=left,
        y axis line style={opacity=0},
        tickwidth=0pt,
        enlarge x limits=true,
        legend style={
            draw=none,
            at={(0.5,-0.3)},
            anchor=center,
            legend columns=-1,
            /tikz/every even column/.append style={column sep=0.5cm}            
        },
        %legend to name={legend_noise},
        tick label style = {font=\MID}, 
        ylabel={Relative error (\%)},
        ylabel style={at={(-0.06,0.5)}},
        ylabel style={font=\HUGE},
        xlabel={Number of repetitions},
        xlabel style={at={(0.5,-0.03)}},
        xlabel style={font=\HUGE},
        symbolic x coords={
           1, 2, 3, 4, 5},
       xtick=data,
       nodes near coords={}
       %nodes near coords={
       % \pgfmathprintnumber[precision=0]{\pgfplotspointmeta}
       %}
    ]
    
   % --------------------------------------------------------
    % Noise-sensitive
    \addplot [
        only marks, 
        mark=square*, 
        mark size=6,
        color=orange!70!white,
        ] 
      coordinates {
      (1, 38.347280)
      (2, 29.322773)
      (3, 35.881262)
      (4, 35.249435)
      (5, 45.376565)};
    % --------------------------------------------------------

    % --------------------------------------------------------
    % Noise-resilient
    \addplot [
        only marks, 
        mark=*, 
        mark size=6,
        blue!70!white
        ] 
      coordinates {
      (1, 31.673943)
      (2, 26.305146)
      (3, 34.148846)
      (4, 30.078374)
      (5, 35.301943)};
    % --------------------------------------------------------

    % --------------------------------------------------------
    % Noise-resilient
    \addplot [
        color=blue!70!white,          
        dashed,             
        line width=1pt     
        ]   
      coordinates {
      (1, 31.673943)
      (2, 26.305146)
      (3, 34.148846)
      (4, 30.078374)
      (5, 35.301943)};
    % --------------------------------------------------------

    % --------------------------------------------------------
    % Noise-sensitive
    \addplot [
        color=orange!70!white,         
        dashed,             
        line width=1pt     
        ]   
      coordinates {
      (1, 38.347280)
      (2, 29.322773)
      (3, 35.881262)
      (4, 35.249435)
      (5, 45.376565)};
    % --------------------------------------------------------
    
  \end{axis}
\end{tikzpicture}

				
    }
    \caption{Benchmark--computation}\label{fig:reduction_bg_comp_RE}
  \end{subfigure} 
 \medskip 
\vspace{1ex}
 \begin{subfigure}[t]{0.48\linewidth} 
    \centering
    \resizebox{\columnwidth}{!}{%
        \makeatletter
\newcommand\HUGE{\@setfontsize\Huge{25}{25}}
\newcommand\MID{\@setfontsize\Mid{20}{20}}
\makeatother

\begin{tikzpicture}
  \centering
  \begin{axis}[
        height=8.5cm, width=11cm,
        bar width=1.0cm,
        ymajorgrids, tick align=inside,
        major grid style={draw=gray!20!white},
        enlarge y limits={value=.1,upper},
        ymin=0, ymax=1.2,
        axis x line*=bottom,
        axis y line*=left,
        y axis line style={opacity=0},
        tickwidth=0pt,
        enlarge x limits=true,
        legend style={
            draw=none,
            at={(0.5,-0.3)},
            anchor=center,
            legend columns=2,
            /tikz/every even column/.append style={column sep=0.5cm}            
        },
        tick label style = {font=\MID}, 
        ylabel={Exponent deviation},
        ylabel style={at={(-0.06,0.5)}},
        ylabel style={font=\HUGE},
        xlabel={Number of repetitions},
        xlabel style={at={(0.5,-0.03)}},
        xlabel style={font=\HUGE},
        symbolic x coords={
           1, 2, 3, 4, 5},
       xtick=data,
       nodes near coords={}
       %nodes near coords={
       % \pgfmathprintnumber[precision=0]{\pgfplotspointmeta}
       %}
    ]       
    
    % --------------------------------------------------------
    % Noise-sensitive
    \addplot [
        only marks, 
        mark=square*, 
        mark size=6,
        color=orange!70!white, 
        ] 
      coordinates {
      (1, 0.89)
      (2, 0.98)
      (3, 1.02)
      (4, 1.05)
      (5, 1.10)};     
    % --------------------------------------------------------

    % --------------------------------------------------------
    % Noise-resilient
    \addplot [
        only marks, 
        mark=*, 
        mark size=6,
        blue!70!white
        ] 
      coordinates {
      (1, 0)
      (2, 0)
      (3, 0)
      (4, 0)
      (5, 0)};
    % --------------------------------------------------------

    % --------------------------------------------------------
    % Noise-resilient
    \addplot [
        color=blue!70!white,          
        dashed,             
        line width=1pt     
        ]   
      coordinates {
      (1, 0)
      (2, 0)
      (3, 0)
      (4, 0)
      (5, 0)};
    % --------------------------------------------------------

    % --------------------------------------------------------
    % Noise-sensitive
    \addplot [
        color=orange!70!white,         
        dashed,             
        line width=1pt     
        ]   
      coordinates {
      (1, 0.89)
      (2, 0.98)
      (3, 1.02)
      (4, 1.05)
      (5, 1.10)}; 
    % --------------------------------------------------------
   
  \end{axis}
\end{tikzpicture}
    }
    \caption{Benchmark--communication}\label{fig:reduction_bg_mpi_ED}
  \end{subfigure} 
\hfill
 \begin{subfigure}[t]{0.48\linewidth} 
    \centering
    \resizebox{\columnwidth}{!}{%
        \makeatletter
\newcommand\HUGE{\@setfontsize\Huge{25}{25}}
\newcommand\MID{\@setfontsize\Mid{20}{20}}
\makeatother

\begin{tikzpicture}
  \centering
  \begin{axis}[
        height=8.5cm, width=11cm,
        bar width=1.0cm,
        ymajorgrids, tick align=inside,
        major grid style={draw=gray!20!white},
        enlarge y limits={value=.1,upper},
        ymin=0, ymax=120,
        axis x line*=bottom,
        axis y line*=left,
        y axis line style={opacity=0},
        tickwidth=0pt,
        enlarge x limits=true,
        legend style={
            draw=none,
            at={(0.5,-0.3)},
            anchor=center,
            legend columns=-1,
            /tikz/every even column/.append style={column sep=0.5cm}            
        },
        %legend to name={legend_noise},
        tick label style = {font=\MID}, 
        ylabel={Relative error (\%)},
        ylabel style={at={(-0.06,0.5)}},
        ylabel style={font=\HUGE},
        xlabel={Number of repetitions},
        xlabel style={at={(0.5,-0.03)}},
        xlabel style={font=\HUGE},
        symbolic x coords={
           1, 2, 3, 4, 5},
       xtick=data,
       nodes near coords={}
       %nodes near coords={
       % \pgfmathprintnumber[precision=0]{\pgfplotspointmeta}
       %}
    ]
    
   % --------------------------------------------------------
    % Noise-sensitive
    \addplot [
        only marks, 
        mark=square*, 
        mark size=6,
        color=orange!70!white, 
        ] 
      coordinates {
      (1, 114.467357)
      (2, 89.019009)
      (3, 98.025821)
      (4, 84.669132)
      (5, 90.567751)};
    % --------------------------------------------------------

    % --------------------------------------------------------
    % Noise-resilient
    \addplot [
        only marks, 
        mark=*, 
        mark size=6,
        blue!70!white
        ] 
      coordinates {
      (1, 75.643342)
      (2, 78.348946)
      (3, 77.366565)
      (4, 78.980797)
      (5, 78.251652)};
    % --------------------------------------------------------

    % --------------------------------------------------------
    % Noise-resilient
    \addplot [
        color=blue!70!white,          
        dashed,             
        line width=1pt     
        ]   
      coordinates {
      (1, 75.643342)
      (2, 78.348946)
      (3, 77.366565)
      (4, 78.980797)
      (5, 78.251652)};
    % --------------------------------------------------------

    % --------------------------------------------------------
    % Noise-sensitive
    \addplot [
        color=orange!70!white,          
        dashed,             
        line width=1pt     
        ]   
      coordinates {
      (1, 114.467357)
      (2, 89.019009)
      (3, 98.025821)
      (4, 84.669132)
      (5, 90.567751)};
    % --------------------------------------------------------
    
  \end{axis}
\end{tikzpicture}

				
    }
    \caption{Benchmark--communication}\label{fig:reduction_bg_mpi_RE}
  \end{subfigure} 
  \ref{legend_reduction_bg}
  \caption{Effect of repeating measurements on the average deviation of our evaluation metrics when comparing our SWC-based approach to classic modeling (i.e., without noise-resilient priors) for the generated benchmarks. We assess the exponent deviation of the performance models and the relative error between predicted and measured values at the test point using models based on varying numbers of repeated measurements.}
  \label{fig:bg_reduction}
\end{figure}

% ----------------------------------------------------------
% Real applications 
% ----------------------------------------------------------

\subsection{Application case studies} \label{sec:eval:real}

We evaluate our approach using two HPC applications. 
\textbf{(i)~Kripke}~\cite{kunen2015kripke} is a mini-app that simulates particle transport via angular fluxes. The problem size is determined by the number of energy groups $G$ and zones $Z$ in a 3D mesh. \textbf{(ii)~RELeARN}~\cite{rinke2018scalable} simulates the rewiring of connections between neurons, with the problem size per rank represented by $n$. We consider weak-style scaling scenarios for both applications with the number of MPI ranks denoted by $p$. \Cref{table:app_codes} summarizes all execution configurations.

\begin{table}
\centering
\caption{The execution configurations used for each application, including training and test points.}
\resizebox{\columnwidth}{!}{\begin{tabular}{l l l} 
\toprule
\textbf{App/System} & \textbf{Training data} & \textbf{Test point}  \\  
\midrule
% Kripke 
\multirow{2}{*}{\begin{tabular}{@{}l@{}} Kripke \\ Lichtenberg II \end{tabular}} 
& \multirow{3}{*}{\begin{tabular}{@{}l@{}} $p \in \{ 512, 1000, 1728, 2744, 4096 \}$ \\ $G \in \{ 32, 64, 96, 128, 160 \}$ \\ $Z \in \{4^3,  8^3, 12^3, 16^3, 20^3\}$ \end{tabular}} 
& \multirow{3}{*}{\begin{tabular}{@{}l@{}} $(p, G, Z) = (5832, 160, 20^3)$ \\ $ (p, G, Z) = (4096, 192, 20^3)$ \\ $(p, G, Z) = (4096, 160, 24^3)$ \end{tabular}}
\\ \\ \\
\midrule
% RELeARN
\multirow{2}{*}{\begin{tabular}{@{}l@{}} RELeARN \\ Jureca-DC \end{tabular}} 
& \multirow{2}{*}{\begin{tabular}{@{}l@{}} $p = \{32, 64,128,256,512\}$ \\ $n = \{250, 300, 350, 400, 450\}$ \end{tabular}} 
& \multirow{2}{*}{\begin{tabular}{@{}l@{}} $(p, n) = (1024, 450)$ \\ $(p, n) = (\ 512, 500)$ \end{tabular}}
\\ \\ 
\bottomrule
\end{tabular}}
\label{table:app_codes}
\end{table}

\subsubsection{Accuracy} \label{sec:eval:real:accuracy}

We analyse the SweepSolver kernel~\cite{bailey2008analysis} in Kripke for computation and communication performance individually. For Kripke-computation, under weak scaling, the parameters $G$ and zones $Z$ scale proportionally with the number of ranks $p$, resulting in a theoretical model of $G \cdot Z$. While for Kripke-communication, the wavefront algorithm induces delays through repeated executions of the MPI\_Testany function. This causes a theoretical scaling model of $p^{1/3}$, which reflects the waiting time growing along the diagonal of the cuboid simulation volume, as messages successively satisfy the data dependencies of the wavefront algorithm that exist between the MPI ranks~\cite{bailey2008analysis}. For RELeARN, we consider the UpdateConnectivity kernel, whose theoretical performance has been defined as $p + n \cdot \log_2(n \cdot p)$~\cite{czappa2023simulating}. 

We begin our analysis by comparing the classic, the DNN-based, and the SWC-based models with their expected theoretical baselines. For each input parameter $x$, we compute the ED to quantify divergence from the theoretical scaling behavior. \Cref{table:cs_model_rq1} presents the deviation values $\Delta x$ per variable. Lower deviations indicate better theoretical alignment. In five out of eight cases, our SWC-based models achieve the closest match to theoretical predictions; in two cases, the results are equivalent to previous models, and in only one case, the performance is slightly worse. 

\begin{table}
\centering
\caption{Performance models for Kripke and RELeARN. The table compares the theoretical complexity of key kernels~\cite{bailey2008analysis,czappa2023simulating} with empirical models derived from classic, DNN-based, and SWC-based approaches. The exponent deviation ($\Delta x$) quantifies the discrepancy between each model’s scaling behavior and the theoretical expectation for parameter $x$. Lower values indicate better alignment; bold entries denote the closest match.}
\resizebox{\columnwidth}{!}{
\begin{tabular}{lp{4.3cm} l l l}
\toprule
\textbf{Model} & \textbf{Asymptotic complexity} & \multicolumn{3}{l}{\textbf{ED}} \\
\midrule
\multicolumn{2}{l}{\textbf{Kripke--computation}} & $\Delta p$ & $\Delta G$ & $\Delta Z$ \\
Theoretical & $\mathcal{O}( G \cdot Z )$ &  &  &  \\
Classic & $\mathcal{O}(p \cdot \log_2^{2}(p) \cdot G^{\frac{3}{4}} \cdot \log_2(G) \cdot Z^{\frac{4}{5}} )$ & $1$ & $0.25$ & $\mathbf{0.20}$ \\
DNN-based & $\mathcal{O}(p \cdot G^{\frac{5}{4}} \cdot Z^{\frac{2}{3}} )$ & $1$ & $0.25$ & $0.33$ \\
SWC-based & $\mathcal{O}(G \cdot \log_2(G) \cdot Z^{\frac{3}{4}})$ & $\mathbf{0}$ & $\mathbf{0}$ & $0.25$ \\ 
\midrule 
\multicolumn{2}{l}{\textbf{Kripke--communication}} & $\Delta p$ & $\Delta G$ & $\Delta Z$ \\
Theoretical & $\mathcal{O}( p^{\frac{1}{3}} + G \cdot Z^{\frac{2}{3}} )$ &  &  &  \\
Classic & $\mathcal{O}(p^{\frac{4}{3}} \cdot \log_2(p) \cdot G^{\frac{3}{4}} \cdot \log_2(G) \cdot Z^{\frac{1}{3}} \cdot \log_2^{2}(Z) )$ & $1$ & $0.25$ & $0.33$ \\
DNN-based   & $\mathcal{O}( G^{\frac{5}{4}} \cdot Z^{\frac{1}{2}} )$ & $\mathbf{0.33}$ & $0.25$ & $0.16$ \\
SWC-based & $\mathcal{O}(G \cdot Z^{\frac{2}{3}} )$ & $\mathbf{0.33}$ & $\mathbf{0}$ & $\mathbf{0}$ \\
\midrule
% Line
\multicolumn{2}{l}{\textbf{RELeARN}} & $\Delta p$ & $\Delta n$ &  \\
Theoretical & $\mathcal{O}( p + n \cdot \log_2(n \cdot p) )$ &  &  &  \\
Classic & $\mathcal{O}( p^{\frac{2}{3}} \cdot n^{\frac{3}{4}} \cdot \log_2(n) )$ & $0.33$ & $\mathbf{0.25}$ & \\ 
DNN-based   & $\mathcal{O}( p^{\frac{2}{3}} \cdot \log_2(p) \cdot n^{\frac{1}{4}} )$ & $0.33$ & $0.75$ & \\ 
SWC-based & $\mathcal{O}(p + n^{\frac{5}{4}} \cdot \log_2(n) \cdot p^{\frac{1}{4}} )$ & $\mathbf{0}$ & $\mathbf{0.25}$ & \\  
\bottomrule
\end{tabular}
}
\label{table:cs_model_rq1}
\end{table}

The theoretical expectation of the waiting time scaling, such as the $p^{1/3}$ term for Kripke-communication,  represents delays associated with data dependencies in the wavefront algorithm. Our SWC-based approach is derived from actual communication effort and does not capture waiting time, resulting in a constant scaling for $p$. Thus, deviations between SWC-based and runtime models may offer insight into wavefront-induced inefficiencies. In our experiment, the difference comes from a mixture of waiting time and noise, which is technically challenging to distinguish. 

We further assess predictive power via RE. \Cref{fig:model_perf} shows the median of both training and test points across all models, and compares the RE values of each model on test points (\cref{table:app_codes}). Our SWC-based models outperform the others in six out of eight cases. Our results show that RE ranges from $2\%$ to $46\%$ for Kripke, while RELeARN achieves a value as low as $3\%$. As in the previous analysis, all models are sufficiently accurate, even in cases with errors exceeding $40\%$, as the predictions still represent the correct order of magnitude---enough for performance diagnosis and bottleneck identification.  

Whiskers in the test point figures represent min--max run-to-run variability. In some cases (\cref{fig:model_1_1,fig:model_1_2,fig:model_3_1,fig:model_3_2}), the range between the maximum and minimum values is not significant---between $0.8\%$ and $19\%$. However, in other experiments (\cref{fig:model_1_3,fig:model_2_1,fig:model_2_2,fig:model_2_3}), we observe substantial variability---between $24\%$ and $38\%$---highlighting the impact of noise on the measurements, especially within communication functions. Additionally, the theoretical models show deviations from the measured data. This discrepancy arises because the models are fitted using the whole training dataset across all parameters, whereas the figures isolate the impact of individual parameters by varying one at a time while holding others constant.

\begin{figure*}[tp!]
\centering
  \begin{subfigure}[t]{0.32\textwidth} 
    \centering
    \resizebox{\columnwidth}{!}{%
        % Define a larger font size
\makeatletter
\newcommand\HUGE{\@setfontsize\Huge{25}{25}}
\newcommand\MID{\@setfontsize\Mid{20}{20}}
\makeatother

\begin{tikzpicture}
    \begin{axis}[
        height=8.5cm, width=17.5cm,        
        grid=major,
        ymajorgrids, tick align=inside,
        major grid style={draw=gray!20!white},
        enlarge y limits={value=.1,upper},
        ymin=0, ymax=16,
        xmin=0,
        axis x line*=bottom,
        axis y line*=left,
        y axis line style={opacity=1},
        tickwidth=0pt,
        legend style={
            draw=none,
            at={(0,0)},
            anchor=north,
            legend columns=6,
            mark size=6,
            mark options={scale=0.6},
            /tikz/every even column/.append style={column sep=0.5cm}            
        },
        legend to name={legend_performance},
        ylabel={Runtime ($\cdot 10^2$ s) },
        ylabel style={at={(-0.03,0.5)}},
        ylabel style={font=\HUGE},
        xlabel={$p$},
        xlabel style={at={(0.5,-0.03)}},
        xlabel style={font=\HUGE},
        xtick={512, 1000, 1728, 2744, 4096, 5832}, % Define numeric x-ticks        
        tick label style={font=\MID}, % Make axis tick labels larger
        nodes near coords={}
    ]
   
    \addplot [black, thick, dashdotdotted, line width=4pt, domain=0:6000, samples=200]  
    {(15.972290736318973 + 0.0006351160898808409 * 160^(1) * 8000^(1))/100 }; 
    \addlegendentry[color=black]{Theoretical}        

    \addplot [orange!70!white, thick, dashed, dash pattern=on 8pt off 4pt, line width=4pt, domain=0:6000, samples=200]
    {( -2.1141957099907933 + 2.9056462282164693e-09 * x^(1) * log2(x)^(2) * 160^(3/4) * log2(160)^(1) * 8000^(4/5) + 1.0237935621506761e-07 * x^(1) * (log2(x))^(2) * 160^(3/4) * log2(160)^(1) + 0.20350101701593856 * 8000^(4/5) )/100 };
    \addlegendentry[color=black]{Classic}  

    \addplot [purple!70!white, ultra thick, dotted, line width=5pt, domain=0:6000, samples=200] 
    {(-3.604903580045306 + 0.00029573493922064144 * x^(1/3) * 160^(5/4) * 8000^(2/3) )/100 };
    \addlegendentry[color=black]{DNN-based}  

     \addplot [blue!70!white, thick, smooth, line width=4pt, domain=0:6000, samples=200] 
     {(3.427517747344458 + 0.025849847174801124 * 8000^(3/4) + 0.0007845815406218731 * 160 * log2(160) * 8000^(3/4))/100};
     \addlegendentry[color=black]{SWC-based}

    %\node[anchor=west, black, font=\HUGE] at (axis cs:100, 15.0) {Relative error (\%)};
    %\node[anchor=west, blue,  font=\HUGE] at (axis cs:100, 13.0) {R = $15.20$};
    %\node[anchor=west, red,   font=\HUGE] at (axis cs:100, 11.0) {NS = $50.60$};    

    \addplot[
            only marks, % Disable line, keep marks
            black, % Marker color
            mark=square*, % Marker style
            mark size=4pt % Marker size
        ] 
    coordinates {
        (512, 7.481642)  
        (1000, 4.623286)  
        (1728, 9.781170)  
        (2744, 9.870470)  
        (4096, 9.813350)
        };  
    \addlegendentry[color=black]{Training data}

    \addplot[
        only marks,
        blue,
        mark=*,
        mark options={draw=blue, fill=white, thick},
        mark size=4pt,
        error bars/.cd,
        y dir=both, y explicit
    ] 
    coordinates {
        % Asymmetric error bars
        (5832, 9.75822) += (0.855119, 0.855119)
    };
    \addlegendentry[color=black]{Test data}

    \addplot[
        only marks,
        blue,
        mark=*,
        mark options={draw=blue, fill=white, thick},
        mark size=6pt,
        error bars/.cd,
        y dir=both, y explicit
    ] 
    coordinates {
        % Asymmetric error bars
        (5832, 9.75822) -= (1.20520, 1.20520) 
    };    

    \end{axis}
\end{tikzpicture}
    }
    \caption{Kripke--computation: varying $p$, \\ $G = 160$, $Z = 8000$.}\label{fig:model_1_1}
  \end{subfigure}  
 \hfill
  \begin{subfigure}[t]{0.32\textwidth} 
    \centering
    \resizebox{\columnwidth}{!}{%
        % Define a larger font size
\makeatletter
\newcommand\HUGE{\@setfontsize\Huge{25}{25}}
\newcommand\MID{\@setfontsize\Mid{20}{20}}
\makeatother

\begin{tikzpicture}
    \begin{axis}[
        height=8.5cm, width=17.5cm,        
        grid=major,
        ymajorgrids, tick align=inside,
        major grid style={draw=gray!20!white},
        enlarge y limits={value=.1,upper},
        ymin=0, ymax=16,
        xmin=0,
        axis x line*=bottom,
        axis y line*=left,
        y axis line style={opacity=1},
        tickwidth=0pt,
        legend style={
            draw=none,
            at={(0,0)},
            anchor=north,
            legend columns=1,
            /tikz/every even column/.append style={column sep=0.5cm}            
        },
        %legend to name={legend_test},
        %
        ylabel={Runtime ($\cdot 10^2$ s) },
        ylabel style={at={(-0.03,0.5)}},
        ylabel style={font=\HUGE},
        xlabel={$G$},
        xlabel style={at={(0.5,-0.03)}},
        xlabel style={font=\HUGE},
        xtick={32, 64, 96, 128, 160, 192}, % Define numeric x-ticks        
        tick label style={font=\MID}, % Make axis tick labels larger
        nodes near coords={}
    ]

    \addplot [black, thick, dashdotdotted, line width=4pt, domain=0:195, samples=200]
    {(15.972290736318973 + 0.0006351160898808409 * x^(1) * 8000^(1))/100}; 
    %\addlegendentry{Theoretical model}
    
    \addplot [orange!70!white, thick, dashed, dash pattern=on 8pt off 4pt, line width=4pt, domain=0:195, samples=200] 
    {( -2.1141957099907933 + 2.9056462282164693e-09 * 4096^(1) * (log2(4096))^(2) * x^(3/4) * log2(x)^(1) * 8000^(4/5) + 1.0237935621506761e-07 * 4096^(1) * (log2(4096)^(2)) * x^(3/4) * log2(x)^(1) + 0.20350101701593856 * 8000^(4/5) )/100};
    %\addlegendentry{Noise-sensitive model} 

    \addplot [purple!70!white, ultra thick, dotted, line width=5pt, domain=0:195, samples=200] 
    {(-3.604903580045306 + 0.00029573493922064144 * 4096^(1/3) * x^(5/4) * 8000^(2/3))/100}; 
    %\addlegendentry{DNN}

     \addplot [blue!70!white, thick, smooth, line width=4pt, domain=0:195, samples=200]  
    {( 3.427517747344458 + 0.025849847174801124 * 8000^(3/4) + 0.0007845815406218731 * x * log2(x) * 8000^(3/4) )/100};
     %\addlegendentry{Noise-resilient model}

    %\node[anchor=west, black, font=\HUGE] at (axis cs:5, 15.0) {Relative error (\%)};
    %\node[anchor=west, blue,  font=\HUGE] at (axis cs:5, 13.0) {R = $21.89$};
    %\node[anchor=west, red,   font=\HUGE] at (axis cs:5, 11.0) {NS = $15.72$};

     \addplot[
            only marks, % Disable line, keep marks
            black, % Marker color
            mark=square*, % Marker style
            mark size=4pt % Marker size
        ] 
    coordinates {
      (32, 2.147542)  
      (64, 3.785197)  
      (96, 5.660627)  
      (128, 7.481922)  
      (160, 9.813350)    
    };
    %\addlegendentry{Training data}

    \addplot[
        only marks,
        blue,
        mark=*,
        mark options={draw=blue, fill=white, thick},
        mark size=6pt,
        error bars/.cd,
        y dir=both, y explicit
    ] 
    coordinates {
        % Asymmetric error bars
        (192, 14.0052) += (1.12428, 1.12428)
    };
    
    \addplot[
        only marks,
        blue,
        mark=*,
        mark options={draw=blue, fill=white, thick},
        mark size=6pt,
        error bars/.cd,
        y dir=both, y explicit
    ] 
    coordinates {
        % Asymmetric error bars
        (192, 14.0052) -= (1.64127, 1.64127) 
    };                    

    \end{axis}
\end{tikzpicture}
    }
    \caption{Kripke--computation: varying $G$, \\ $p = 4096$, $Z = 8000$.}\label{fig:model_1_2}
  \end{subfigure}
 \hfill
  \begin{subfigure}[t]{0.32\textwidth} 
    \centering
    \resizebox{\columnwidth}{!}{%
        % Define a larger font size
\makeatletter
\newcommand\HUGE{\@setfontsize\Huge{25}{25}}
\newcommand\MID{\@setfontsize\Mid{20}{20}}
\makeatother

\begin{tikzpicture}
    \begin{axis}[
        height=8.5cm, width=17.5cm,        
        grid=major,
        ymajorgrids, tick align=inside,
        major grid style={draw=gray!20!white},
        enlarge y limits={value=.1,upper},
        ymin=0, ymax=16, %ymax=5.6,
        xmin=0,
        axis x line*=bottom,
        axis y line*=left,
        y axis line style={opacity=1},
        tickwidth=0pt,
        legend style={
            draw=none,
            at={(0,0)},
            anchor=north,
            legend columns=1,
            /tikz/every even column/.append style={column sep=0.5cm}            
        },
        %legend to name={legend_test},
        %
        ylabel={Runtime ($\cdot 10^2$ s) },
        ylabel style={at={(-0.03,0.5)}},
        ylabel style={font=\HUGE},
        xlabel={$Z$},
        xlabel style={at={(0.5,-0.03)}},
        xlabel style={font=\HUGE},
        xtick={1728, 4096, 8000, 13824}, % Define numeric x-ticks        
        tick label style={font=\MID}, % Make axis tick labels larger
        nodes near coords={}
    ]

    \addplot [black, thick, dashdotdotted, line width=4pt, domain=0:14000, samples=200] 
    {(15.972290736318973 + 0.0006351160898808409 * 160^(1) * x^(1))/100}; 
    %\addlegendentry{Theoretical model}
   
    \addplot [orange!70!white, thick, dashed, dash pattern=on 8pt off 4pt, line width=4pt, domain=0:14000, samples=200] 
    {( -2.1141957099907933 + 2.9056462282164693e-09 * 4096^(1) * (log2(4096))^(2) * 160^(3/4) * log2(160)^(1) * x^(4/5) + 1.0237935621506761e-07 * 4096^(1) * (log2(4096)^(2)) * 160^(3/4) * log2(160)^(1) + 0.20350101701593856 * x^(4/5) )/100};
    %\addlegendentry{Noise-sensitive model}    

     \addplot [purple!70!white, ultra thick, dotted, line width=5pt, domain=0:14000, samples=200] 
    {(-3.604903580045306 + 0.00029573493922064144 * 4096^(1/3) * 160^(5/4) * x^(2/3))/100}; 
    %\addlegendentry{DNN}

     \addplot [blue!70!white, thick, smooth, line width=4pt, domain=0:14000, samples=200]  
    {( 3.427517747344458 + 0.025849847174801124 * x^(3/4) + 0.0007845815406218731 * 160 * log2(160) * x^(3/4) )/100};
     %\addlegendentry{Noise-resilient model}

    %\node[anchor=west, black, font=\HUGE] at (axis cs:300, 15.0) {Relative error (\%)};
    %\node[anchor=west, blue,  font=\HUGE] at (axis cs:300, 13.0) {NR = $2.76$};
    %\node[anchor=west, red,   font=\HUGE] at (axis cs:300, 11.0) {NS = $28.35$};   

     \addplot[
            only marks, % Disable line, keep marks
            black, % Marker color
            mark=square*, % Marker style
            mark size=4pt % Marker size
        ] 
    coordinates {
    (64, 0.120502)  
    (512, 0.402353)  
    (1728, 2.970650)  
    (4096, 5.935020)  
    (8000, 9.813350)
    };
    %\addlegendentry{Training data}   
   
    \addplot[
        only marks,
        blue,
        mark=*,
        mark options={draw=blue, fill=white, thick},
        mark size=6pt,
        error bars/.cd,
        y dir=both, y explicit
    ] 
    coordinates {
        % Asymmetric error bars
        (13824, 12.42572) += (3.59821, 3.59821)
    };
    
    \addplot[
        only marks,
        blue,
        mark=*,
        mark options={draw=blue, fill=white, thick},
        mark size=6pt,
        error bars/.cd,
        y dir=both, y explicit
    ] 
    coordinates {
        % Asymmetric error bars
        (13824, 12.42572) -= (1.62225, 1.62225) 
    };     
    
    \end{axis}
\end{tikzpicture}
    }
    \caption{Kripke--computation: varying $Z$, \\ $p = 4096$, $G = 160$.}\label{fig:model_1_3}
  \end{subfigure}  
  \medskip    
  \begin{subfigure}[t]{0.32\textwidth} 
    \centering
    \resizebox{\columnwidth}{!}{%
        % Define a larger font size
\makeatletter
\newcommand\HUGE{\@setfontsize\Huge{25}{25}}
\newcommand\MID{\@setfontsize\Mid{20}{20}}
\makeatother

\begin{tikzpicture}
    \begin{axis}[
        height=8.5cm, width=17.5cm,        
        grid=major,
        ymajorgrids, tick align=inside,
        major grid style={draw=gray!20!white},
        enlarge y limits={value=.1,upper},
        ymin=0, ymax=16,
        xmin=0,
        axis x line*=bottom,
        axis y line*=left,
        y axis line style={opacity=1},
        tickwidth=0pt,
        legend style={
            draw=none,
            at={(0,0)},
            anchor=north,
            legend columns=5,
            mark size=6,
            mark options={scale=0.6},
            /tikz/every even column/.append style={column sep=0.5cm}            
        },
        legend to name={legend_test},
        ylabel={Runtime ($\cdot 10^2$ s) },
        ylabel style={at={(-0.03,0.5)}},
        ylabel style={font=\HUGE},
        xlabel={$p$},
        xlabel style={at={(0.5,-0.03)}},
        xlabel style={font=\HUGE},
        xtick={512, 1000, 1728, 2744, 4096, 5832}, % Define numeric x-ticks        
        tick label style={font=\MID}, % Make axis tick labels larger
        nodes near coords={}
    ]

    \addplot [black, thick, dashdotdotted, line width=4pt, domain=0:6000, samples=200] 
    {(-231.54085971586642 + 0.009862956702543595 * 160^(1) * 8000^(2/3) + 19.376878927606146 * x^(1/3))/100}; 
    %\addlegendentry{Theoretical model}   

    \addplot [orange!70!white, thick, dashed, dash pattern=on 8pt off 4pt, line width=4pt, domain=0:6000, samples=200] 
    {( -9.447386453461554 + 6.077190970125065e-10 * x^(4/3) * log2(x)^(1) * 160^(3/4) * log2(160)^(1) * 8000^(1/3) * (log2(8000))^(2) + 0.00037114803946612163 * 160^(3/4) * log2(160)^(1) * 8000^(1/3) * (log2(8000))^(2) )/100};
    %\addlegendentry{Noise-sensitive model} 

    \addplot [purple!70!white, ultra thick, dotted, line width=5pt, domain=0:6000, samples=200] 
    {(-7.712281000671503 + 0.01043983048620487 * 160^(5/4) * 8000^(1/2))/100}; 
    %\addlegendentry{DNN}  

    \addplot [blue!70!white, thick, smooth, line width=4pt, domain=0:6000, samples=200]  
    {( 9.651651147720026 + 0.009163846772279417 * 160 * 8000^(2/3) )/100};
    %\addlegendentry{Noise-resilient model}

    %\node[anchor=west, black, font=\HUGE] at (axis cs:100, 15.0) {Relative error (\%)};
    %\node[anchor=west, blue,  font=\HUGE] at (axis cs:100, 13.0) {R = $26.78$};
    %\node[anchor=west, red,   font=\HUGE] at (axis cs:100, 11.0) {NS = $57.88$};

    \addplot[
            only marks, % Disable line, keep marks
            black, % Marker color
            mark=square*, % Marker style
            mark size=4pt % Marker size
        ] 
    coordinates {
        (512, 3.795700)  
        (1000, 4.020270)  
        (1728, 7.467820)  
        (2744, 8.670090)  
        (4096, 8.258300) 
        };  
    %\addlegendentry{Training data}  

    \addplot[
        only marks,
        blue,
        mark=*,
        mark options={draw=blue, fill=white, thick},
        mark size=4pt,
        error bars/.cd,
        y dir=both, y explicit
    ] 
    coordinates {
        % Asymmetric error bars
        (5832, 8.14200) += (1.49083, 1.49083)
    };

    \addplot[
        only marks,
        blue,
        mark=*,
        mark options={draw=blue, fill=white, thick},
        mark size=6pt,
        error bars/.cd,
        y dir=both, y explicit
    ] 
    coordinates {
        % Asymmetric error bars
        (5832, 8.14200) -= (1.65013, 1.65013) 
    };

    \end{axis}
\end{tikzpicture}
    }
    \caption{Kripke--communication: varying $p$, \\ $G = 160$, $Z = 8000$.}\label{fig:model_2_1}
  \end{subfigure}  
 \hfill
  \begin{subfigure}[t]{0.32\textwidth} 
    \centering
    \resizebox{\columnwidth}{!}{%
        % Define a larger font size
\makeatletter
\newcommand\HUGE{\@setfontsize\Huge{25}{25}}
\newcommand\MID{\@setfontsize\Mid{20}{20}}
\makeatother

\begin{tikzpicture}
    \begin{axis}[
        height=8.5cm, width=17.5cm,        
        grid=major,
        ymajorgrids, tick align=inside,
        major grid style={draw=gray!20!white},
        enlarge y limits={value=.1,upper},
        ymin=0, ymax=16,
        xmin=0,
        axis x line*=bottom,
        axis y line*=left,
        y axis line style={opacity=1},
        tickwidth=0pt,
        legend style={
            draw=none,
            at={(0,0)},
            anchor=north,
            legend columns=1,
            /tikz/every even column/.append style={column sep=0.5cm}            
        },
        %legend to name={legend_test},
        %
        ylabel={Runtime ($\cdot 10^2$ s) },
        ylabel style={at={(-0.03,0.5)}},
        ylabel style={font=\HUGE},
        xlabel={$G$},
        xlabel style={at={(0.5,-0.03)}},
        xlabel style={font=\HUGE},
        xtick={32, 64, 96, 128, 160, 192}, % Define numeric x-ticks        
        tick label style={font=\MID}, % Make axis tick labels larger
        nodes near coords={}
    ]

    \addplot [black, thick, dashdotdotted, line width=4pt, domain=0:195, samples=200]
    {(-231.54085971586642 + 0.009862956702543595 * x^(1) * 8000^(2/3) + 19.376878927606146 * 4096^(1/3))/100}; 
    %\addlegendentry{Theoretical model}

    \addplot [orange!70!white, thick, dashed, dash pattern=on 8pt off 4pt, line width=4pt, domain=0:195, samples=200] 
    {( -9.447386453461554 + 6.077190970125065e-10 * 4096^(4/3) * log2(4096)^(1) * x^(3/4) * log2(x)^(1) * 8000^(1/3) * (log2(8000))^(2) + 0.00037114803946612163 * x^(3/4) * log2(x)^(1) * 8000^(1/3) * (log2(8000)^(2)) )/100};
    %\addlegendentry{Noise-sensitive model}     

    \addplot [purple!70!white, ultra thick, dotted, line width=5pt, domain=0:195, samples=200] 
    {(-7.712281000671503 + 0.01043983048620487 * x^(5/4) * 8000^(1/2))/100}; 
    %\addlegendentry{DNN}   

    \addplot [blue!70!white, thick, smooth, line width=4pt, domain=0:6000, domain=0:195, samples=200] 
    {( 9.651651147720026 + 0.009163846772279417 * x * 8000^(2/3) )/100};
    %\addlegendentry{Noise-resilient model}

    %\node[anchor=west, black, font=\HUGE] at (axis cs:5, 15.0) {Relative error (\%)};
    %\node[anchor=west, blue,  font=\HUGE] at (axis cs:5, 13.0) {NR = $46.07$};
    %\node[anchor=west, red,   font=\HUGE] at (axis cs:5, 11.0) {NS = $16.28$};

     \addplot[
            only marks, % Disable line, keep marks
            black, % Marker color
            mark=square*, % Marker style
            mark size=4pt % Marker size
        ] 
    coordinates {
        (32, 2.068130)  
        (64, 3.203490)  
        (96, 4.826690)  
        (128, 6.748480)  
        (160, 8.258300) 
    };
    %\addlegendentry{Training data}

    \addplot[
        only marks,
        blue,
        mark=*,
        mark options={draw=blue, fill=white, thick},
        mark size=6pt,
        error bars/.cd,
        y dir=both, y explicit
    ] 
    coordinates {
        % Asymmetric error bars
        (192, 13.2290) += (1.6930, 1.6930)
    };

    \addplot[
        only marks,
        blue,
        mark=*,
        mark options={draw=blue, fill=white, thick},
        mark size=6pt,
        error bars/.cd,
        y dir=both, y explicit
    ] 
    coordinates {
        % Asymmetric error bars
        (192, 13.2290) -= (1.8940, 1.8940) 
    };

    \end{axis}
\end{tikzpicture}
    }
    \caption{Kripke--communication: varying $G$, \\ $p = 4096$, $Z = 8000$.}\label{fig:model_2_2}
  \end{subfigure}
 \hfill
  \begin{subfigure}[t]{0.32\textwidth} 
    \centering
    \resizebox{\columnwidth}{!}{%
        % Define a larger font size
\makeatletter
\newcommand\HUGE{\@setfontsize\Huge{25}{25}}
\newcommand\MID{\@setfontsize\Mid{20}{20}}
\makeatother

\begin{tikzpicture}
    \begin{axis}[
        height=8.5cm, width=17.5cm,        
        grid=major,
        ymajorgrids, tick align=inside,
        major grid style={draw=gray!20!white},
        enlarge y limits={value=.1,upper},
        ymin=0, ymax=16, %ymax=5.6,
        xmin=0,
        axis x line*=bottom,
        axis y line*=left,
        y axis line style={opacity=1},
        tickwidth=0pt,
        legend style={
            draw=none,
            at={(0,0)},
            anchor=north,
            legend columns=1,
            /tikz/every even column/.append style={column sep=0.5cm}            
        },
        %legend to name={legend_test},
        %
        ylabel={Runtime ($\cdot 10^2$ s) },
        ylabel style={at={(-0.03,0.5)}},
        ylabel style={font=\HUGE},
        xlabel={$Z$},
        xlabel style={at={(0.5,-0.03)}},
        xlabel style={font=\HUGE},
        xtick={1728, 4096, 8000, 13824}, % Define numeric x-ticks        
        tick label style={font=\MID}, % Make axis tick labels larger
        nodes near coords={}
    ]

    \addplot [black, thick, dashdotdotted, line width=4pt, domain=0:14000, samples=200]
    {(-231.54085971586642 + 0.009862956702543595 * 160^(1) * x^(2/3) + 19.376878927606146 * 4096^(1/3))/100}; 
    %\addlegendentry{Theoretical model}

    \addplot [orange!70!white, thick, dashed, dash pattern=on 8pt off 4pt, line width=4pt, domain=0:14000, samples=200]
    {( -9.447386453461554 + 6.077190970125065e-10 * 4096^(4/3) * log2(4096)^(1) * 160^(3/4) * log2(160)^(1) * x^(1/3) * (log2(x))^(2) + 0.00037114803946612163 * 160^(3/4) * log2(160)^(1) * x^(1/3) * (log2(x))^(2) )/100};
    %\addlegendentry{Noise-sensitive model}    

   \addplot [purple!70!white, ultra thick, dotted, line width=5pt, domain=0:14000, samples=200] 
    {(-7.712281000671503 + 0.01043983048620487 * 160^(5/4) * x^(1/2))/100}; 
    %\addlegendentry{DNN}

    \addplot [blue!70!white, thick, smooth, line width=4pt, domain=0:14000, samples=200]
    {( 9.651651147720026 + 0.009163846772279417 * 160 * x^(2/3) )/100};
    %\addlegendentry{Noise-resilient model}

    %\node[anchor=west, black, font=\HUGE] at (axis cs:300, 15.0) {Relative error (\%)};
    %\node[anchor=west, blue,  font=\HUGE] at (axis cs:300, 13.0) {NR = $18.46$};
    %\node[anchor=west, red,   font=\HUGE] at (axis cs:300, 11.0) {NS = $20.31$};

     \addplot[
            only marks, % Disable line, keep marks
            black, % Marker color
            mark=square*, % Marker style
            mark size=4pt % Marker size
        ] 
    coordinates {
    (64, 0.124276)  
    (512, 0.361315)  
    (1728, 3.654790)  
    (4096, 6.156840)  
    (8000, 8.258300)
    };
    %\addlegendentry{Training data}

    \addplot[
        only marks,
        blue,
        mark=*,
        mark options={draw=blue, fill=white, thick},
        mark size=6pt,
        error bars/.cd,
        y dir=both, y explicit
    ] 
    coordinates {
        % Asymmetric error bars
        (13824, 10.47690) += (4.59260, 4.59260)
    };

    \addplot[
        only marks,
        blue,
        mark=*,
        mark options={draw=blue, fill=white, thick},
        mark size=6pt,
        error bars/.cd,
        y dir=both, y explicit
    ] 
    coordinates {
        % Asymmetric error bars
        (13824, 10.47690) -= (1.23676, 1.23676) 
    };
         
    \end{axis}
\end{tikzpicture}
    }
    \caption{Kripke--communication: varying $Z$, \\ $p = 4096$, $G = 160$.}\label{fig:model_2_3}
  \end{subfigure}  
  \medskip    
  \begin{subfigure}[t]{0.32\textwidth} 
    \centering
    \resizebox{\columnwidth}{!}{%
        % Define a larger font size
\makeatletter
\newcommand\HUGE{\@setfontsize\Huge{25}{25}}
\newcommand\MID{\@setfontsize\Mid{20}{20}}
\makeatother

\begin{tikzpicture}
    \begin{axis}[
        height=8.5cm, width=17.5cm,        
        grid=major,
        ymajorgrids, tick align=inside,
        major grid style={draw=gray!20!white},
        enlarge y limits={value=.1,upper},
        ymin=0, ymax=2.5,
        xmin=0,
        axis x line*=bottom,
        axis y line*=left,
        y axis line style={opacity=1},
        tickwidth=0pt,
        legend style={
            draw=none,
            at={(0,0)},
            anchor=north,
            legend columns=1,
            /tikz/every even column/.append style={column sep=0.5cm}            
        },
        %legend to name={legend_test},
        %
        ylabel={Runtime ($\cdot 10^2$ s) },
        ylabel style={at={(-0.03,0.5)}},
        ylabel style={font=\HUGE},
        xlabel={$p$},
        xlabel style={at={(0.5,-0.03)}},
        xlabel style={font=\HUGE},
        xtick={32, 128, 256, 512, 1024}, % Define numeric x-ticks        
        tick label style={font=\MID}, % Make axis tick labels larger
        nodes near coords={}
    ]

    \addplot [black, thick, dashdotdotted, line width=4pt, domain=0:1050, samples=200]
    {(13.2165 * log2(x) + 0.1204  * x - 44.5500) / 100}; 
    %\addlegendentry{Theoretical model}    

    \addplot [orange!70!white, thick, dashed, dash pattern=on 8pt off 4pt, line width=4pt, domain=0:1050, samples=200] 
    {( -6.278849220678224 + -0.30095028931184625 * x^(2/3) + 0.003138094056811702 * x^(2/3) * 450^(3/4) * log2(450)^(1) ) / 100};
    %\addlegendentry{Dense} 

    \addplot [purple!70!white, ultra thick, dotted, line width=5pt, domain=0:1050, samples=200] 
    {(3.4255015320811855 + -2.7866594440343477 * x^(4/5) + 0.8183351250131174 * x^(4/5) * 450^(1/4)) / 100}; 
    %\addlegendentry{DNN}   

     \addplot [blue!70!white, thick, smooth, line width=4pt, domain=0:1050, samples=200]  
    {( 3.7031265777357927 + -0.005569353659240143 * 450^(5/4) * log2(450) + 0.002353373006996428 * x^(1/4) * 450^(5/4) * log2(450) + 3.703152686923925 + 0.05188595707813407 * x ) / 100 };
     %\addlegendentry{Noise-resilient model}    

    %\node[anchor=west, black, font=\HUGE] at (axis cs:15, 2.4) {Relative error (\%)};
    %\node[anchor=west, blue,  font=\HUGE] at (axis cs:15, 2.1) {R = $4.89$};
    %\node[anchor=west, red,   font=\HUGE] at (axis cs:15, 1.8) {B = $10.61$}; 
          
     \addplot[
            only marks, % Disable line, keep marks
            black, % Marker color
            mark=square*, % Marker style
            mark size=4pt % Marker size
        ] 
    coordinates {
      (32,   0.177935)
      (64,   0.273539)
      (128,  0.634788)
      (256,  1.019552)
      (512,  1.389867)          
    };
    %\addlegendentry{Train data}

    \addplot[
        only marks,
        blue,
        mark=*,
        mark options={draw=blue, fill=white, thick},
        mark size=6pt,
        error bars/.cd,
        y dir=both, y explicit
    ] 
    coordinates {
        % Asymmetric error bars
        (1024, 2.1056) += (0.00913500, 0.00913500)
    };
    
    \addplot[
        only marks,
        blue,
        mark=*,
        mark options={draw=blue, fill=white, thick},
        mark size=6pt,
        error bars/.cd,
        y dir=both, y explicit
    ] 
    coordinates {
        % Asymmetric error bars
        (1024, 2.1056) -= (0.00782700, 0.00782700) 
    };

    \end{axis}
\end{tikzpicture}

          
    }
    \caption{RELeARN: varying $p$, $n = 450$.}\label{fig:model_3_1}
  \end{subfigure}
  \hfill
 %\hspace{0.01\textwidth}
  \begin{subfigure}[t]{0.32\textwidth} 
    \centering
    \resizebox{\columnwidth}{!}{%
        % Define a larger font size
\makeatletter
\newcommand\HUGE{\@setfontsize\Huge{25}{25}}
\newcommand\MID{\@setfontsize\Mid{20}{20}}
\makeatother

\begin{tikzpicture}
    \begin{axis}[
        height=8.5cm, width=17.5cm,        
        grid=major,
        ymajorgrids, tick align=inside,
        major grid style={draw=gray!20!white},
        enlarge y limits={value=.1,upper},
        ymin=0, ymax=2.5,
        xmin=0,
        axis x line*=bottom,
        axis y line*=left,
        y axis line style={opacity=1},
        tickwidth=0pt,
        legend style={
            draw=none,
            at={(0,0)},
            anchor=north,
            legend columns=1,
            /tikz/every even column/.append style={column sep=0.5cm}            
        },
        %legend to name={legend_test},
        %
        ylabel={Runtime ($\cdot 10^3$ s) },
        ylabel style={at={(-0.03,0.5)}},
        ylabel style={font=\HUGE},
        xlabel={$n$},
        xlabel style={at={(0.5,-0.03)}},
        xlabel style={font=\HUGE},
        xtick={250, 300, 350, 400, 450, 500}, % Define numeric x-ticks        
        tick label style={font=\MID}, % Make axis tick labels larger
        nodes near coords={}
    ]

    \addplot [black, thick, dashdotdotted, line width=4pt, domain=0:510, samples=200]
    {(0.2643 * x  + 17.0948) / 100}; 
    %\addlegendentry{Theoretical model}

    \addplot [orange!70!white, thick, dashed, dash pattern=on 8pt off 4pt, line width=4pt, domain=0:510, samples=200] 
    {( -6.278849220678224 + -0.30095028931184625 * 512^(2/3) + 0.003138094056811702 * 512^(2/3) * x^(3/4) * log2(x)^(1) ) / 100};
    %\addlegendentry{Noise-sensitive model} 

    \addplot [purple!70!white, ultra thick, dotted, line width=5pt, domain=0:510, samples=200] 
    {(3.4255015320811855 + -2.7866594440343477 * 512^(4/5) + 0.8183351250131174 * 512^(4/5) * x^(1/4)) / 100}; 
    %\addlegendentry{DNN}

     \addplot [blue!70!white, thick, smooth, line width=4pt, domain=0:510, samples=200]  
    {( 3.7031265777357927 + -0.005569353659240143 * x^(5/4) * log2(x) + 0.002353373006996428 * 512^(1/4) * x^(5/4) * log2(x) + 3.703152686923925 + 0.05188595707813407 * 512 ) / 100 };
    % \addlegendentry{Noise-resilient model}

    %\node[anchor=west, black, font=\HUGE] at (axis cs:10, 2.4) {Relative error (\%)};
    %\node[anchor=west, blue,  font=\HUGE] at (axis cs:10, 2.1) {R = $2.21$};
    %\node[anchor=west, red,   font=\HUGE] at (axis cs:10, 1.8) {B = $10.65$};

     \addplot[
            only marks, % Disable line, keep marks
            black, % Marker color
            mark=square*, % Marker style
            mark size=4pt % Marker size
        ] 
    coordinates {
      (250, 0.784096)
      (300, 0.956169)
      (350, 1.078318)
      (400, 1.229960)
      (450, 1.389867)          
    };
    %\addlegendentry{Train data}

    \addplot[
        only marks,
        blue,
        mark=*,
        mark options={draw=blue, fill=white, thick},
        mark size=6pt,
        error bars/.cd,
        y dir=both, y explicit
    ] 
    coordinates {
        % Asymmetric error bars
        (500, 1.4928) += (0.01194100, 0.01194100)
    };
    
    \addplot[
        only marks,
        blue,
        mark=*,
        mark options={draw=blue, fill=white, thick},
        mark size=6pt,
        error bars/.cd,
        y dir=both, y explicit
    ] 
    coordinates {
        % Asymmetric error bars
        (500, 1.4928) -= (0.00143000, 0.00143000) 
    };

    \end{axis}
\end{tikzpicture}

          
    }
    \caption{RELeARN: varying $n$, $p = 512$.}\label{fig:model_3_2}
  \end{subfigure}
  \hfill
  \begin{subfigure}[t]{0.32\textwidth} 
    \centering
    \resizebox{\columnwidth}{!}{%
        % Define a larger font size
\usetikzlibrary{patterns.meta}
\makeatletter
\newcommand\HUGE{\@setfontsize\Huge{25}{25}}
\makeatother

\begin{tikzpicture}
\begin{axis}[
    ybar,
    bar width=8pt,
    width=17.5cm,
    height=8.5cm,
    grid=major,
    ymajorgrids, tick align=inside,
    major grid style={draw=gray!20!white},
    ymin=0,
    ymax=60,
    ylabel={Relative error (\%)},
    ylabel style={at={(-0.03,0.5)}, font=\HUGE},
    xlabel style={at={(0.5,-0.03)}, font=\HUGE},
    symbolic x coords={
        \ref{fig:model_1_1}, \ref{fig:model_1_2}, \ref{fig:model_1_3},
        \ref{fig:model_2_1}, \ref{fig:model_2_2}, \ref{fig:model_2_3},
        \ref{fig:model_3_1}, \ref{fig:model_3_2}
    },
    xtick=data,
    xticklabel style={rotate=45, anchor=east, font=\Huge},
    tick label style={font=\Huge},
    enlarge x limits={abs=0.75cm},
    axis x line*=bottom,
    axis y line*=left,
    tickwidth=0pt,
    nodes near coords={},
    %every node near coord/.append style={font=\small},
    legend style={draw=none},
    cycle list={
        {orange!70!white, fill=orange!70!white},  
        %{draw=purple!80!white, pattern=dots, pattern color=purple!70!white, fill=white, line width=1pt},
        {purple!80!white, fill=purple!70!white},
        {blue!70!white, fill=blue!70!white}
    }
]

% Noise-sensitive
\addplot+[] coordinates {
    (\ref{fig:model_1_1},49)
    (\ref{fig:model_1_2},15)
    (\ref{fig:model_1_3},28)
    (\ref{fig:model_2_1},57)
    (\ref{fig:model_2_2},16)
    (\ref{fig:model_2_3},20)
    (\ref{fig:model_3_1},10)
    (\ref{fig:model_3_2},10)
};

% DNN-based
\addplot+[] coordinates {
    (\ref{fig:model_1_1},23)
    (\ref{fig:model_1_2},3)
    (\ref{fig:model_1_3},24)
    (\ref{fig:model_2_1},35)
    (\ref{fig:model_2_2},50)
    (\ref{fig:model_2_3},34)
    (\ref{fig:model_3_1},21)
    (\ref{fig:model_3_2},9)
};

% Noise-resilient
\addplot+[] coordinates {
    (\ref{fig:model_1_1},17)
    (\ref{fig:model_1_2},21)
    (\ref{fig:model_1_3},2)
    (\ref{fig:model_2_1},26)
    (\ref{fig:model_2_2},46)
    (\ref{fig:model_2_3},18)
    (\ref{fig:model_3_1},3)
    (\ref{fig:model_3_2},3)
};

\end{axis}
\end{tikzpicture} 
    }
    \caption{Relative error values of each model on test points.}\label{fig:per_bar}
  \end{subfigure}
  \ref{legend_performance}
  \caption{Generated performance models are presented alongside the median values of the measured training and test points. The figure illustrates the range of variation at the test point, with the blue whiskers indicating the span between the minimum and maximum execution time. The relative error at the test point is given for the classic, DNN-based, and SWC-based models.}  
  \label{fig:model_perf}
\end{figure*}

\subsubsection{Robustness to noise} \label{sec:eval:real:robust}

Expanding on the previous analysis, we apply artificial noise to the runtime measurements with ascending noise intensities of $2\%$, $5\%$, $10\%$, $50\%$, and $75\%$.
As in the synthetic evaluation, we subject the training data of each function to $100$ independent random noise injections and subsequently create a performance model for each. \Cref{fig:app_noise} shows the mean and standard deviation of the accuracy metrics after applying noise. 

Initially, we aim to validate how the noise affects the asymptotic complexity of the models in \cref{fig:app_kripke1,fig:app_kripke2,fig:app_relearn1}. SWC-based models maintain a consistent mean of exponent deviation across all noise levels: $0.08$ (Kripke--computation), $0.11$ (Kripke--communication), and $0.12$ (RELeARN). In contrast, classic and DNN-based models show significant variance, with ED up to $0.50$. With noise levels up to $10\%$, classic and DNN-based models suffer, with RE values up to $75\%$, $80\%$, and $70\%$ in the most critical scenarios. In comparison, our SWC-based approach limits errors to $35\%$, $20\%$, and $50\%$. These findings confirm our method’s resilience to performance variability.

\begin{figure*}[tp!]
\centering
  \begin{subfigure}[t]{0.32\textwidth} 
    \centering
    \resizebox{\columnwidth}{!}{%
        \makeatletter
\newcommand\HUGE{\@setfontsize\Huge{25}{25}}
\newcommand\MID{\@setfontsize\Mid{20}{20}}
\makeatother

\begin{tikzpicture}
  \centering
  \begin{axis}[
        height=8.5cm, width=11cm,
        bar width=1.0cm,
        ymajorgrids, tick align=inside,
        major grid style={draw=gray!20!white},
        enlarge y limits={value=.1,upper},
        ymin=0, ymax=0.6,
        axis x line*=bottom,
        axis y line*=left,
        y axis line style={opacity=0},
        tickwidth=0pt,
        enlarge x limits=true,
        legend style={
            draw=none,
            at={(0.5,-0.3)},
            anchor=center,
            legend columns=-1,
            mark size=6,
            mark options={scale=0.6},
            /tikz/every even column/.append style={column sep=0.5cm}            
        },
        legend to name={legend_noise},
        tick label style = {font=\MID}, 
        ylabel={Exponent deviation},
        ylabel style={at={(-0.06,0.5)}},
        ylabel style={font=\MID},
        xlabel={Artificial noise},
        xlabel style={at={(0.5,-0.03)}},
        xlabel style={font=\MID},
        symbolic x coords={
           2\%, 5\%, 10\%, 50\%, 75\%},
       xtick=data,
       nodes near coords={}
       %nodes near coords={
       % \pgfmathprintnumber[precision=0]{\pgfplotspointmeta}
       %}
    ]  

    % --------------------------------------------------------
    % Noise-sensitive
    \addplot [
        only marks, 
        mark=square*, 
        mark size=6,
        color=orange!70!white,   
        ] 
      coordinates {
      (2\%, 0.48)
      (5\%, 0.46)
      (10\%, 0.46)
      (50\%, 0.44)
      (75\%, 0.43)};
    \addlegendentry[color=black]{Classic}  
    % --------------------------------------------------------   

    % --------------------------------------------------------
    % DNN-based
    \addplot [
        only marks, 
        mark=triangle*, 
        mark size=6,
        color=purple!80!white,
        ] 
      coordinates {
      (2\%, 0.5277)
      (5\%, 0.5277)
      (10\%, 0.5277)
      (50\%, 0.5291)
      (75\%, 0.52611)};
    \addlegendentry[color=black]{DNN-based}  
    % -------------------------------------------------------- 

    % --------------------------------------------------------
    % Noise-resilient
    \addplot [
        only marks, 
        mark=*, 
        mark size=6,
        blue!70!white
        ]   
      coordinates {
      (2\%, 0.08)
      (5\%, 0.08)
      (10\%, 0.08)
      (50\%, 0.08)
      (75\%, 0.08)};
    \addlegendentry[color=black]{SWC-based}  
    % --------------------------------------------------------
    
    % --------------------------------------------------------
    % Noise-sensitive
    \addplot [
        color=orange!70!white,          
        dashed,             
        line width=1pt     
        ]   
      coordinates {
      (2\%, 0.483) 
      (5\%, 0.469) 
      (10\%, 0.455) 
      (50\%, 0.436) 
      (75\%, 0.447) };
    % Standard deviation region (example values for std)
    \addplot[name path=sensitive_upper, draw=none] 
    coordinates {
      (2\%, 0.491) 
      (5\%, 0.500) 
      (10\%, 0.494) 
      (50\%, 0.572) 
      (75\%, 0.583)};   
    \addplot[name path=sensitive_lower, draw=none] 
    coordinates {
      (2\%, 0.475) 
      (5\%, 0.438) 
      (10\%, 0.416) 
      (50\%, 0.300) 
      (75\%, 0.311)};
    % Fill standard deviation region
    \addplot[color=orange!40!white, opacity=0.5] fill between[of=sensitive_upper and sensitive_lower];
    % --------------------------------------------------------    

    % --------------------------------------------------------
    % DNN-based
    \addplot [
        color=purple!80!white,         
        dashed,             
        line width=1pt     
        ]   
      coordinates {
      (2\%, 0.5277)
      (5\%, 0.5277)
      (10\%, 0.5277)
      (50\%, 0.5291)
      (75\%, 0.52611)};
    % Standard deviation region (example values for std)
    \addplot[name path=dnn_upper, draw=none] 
    coordinates {
      (2\%, 0.5277)
      (5\%, 0.5277)
      (10\%, 0.5277)
      (50\%, 0.5729)
      (75\%,  0.58488)};   
    \addplot[name path=dnn_lower, draw=none] 
    coordinates {
      (2\%, 0.5277)
      (5\%, 0.5277)
      (10\%, 0.5277)
      (50\%, 0.48535)
      (75\%, 0.46734)};
    % Fill standard deviation region
    \addplot[color=purple!40!white, opacity=0.5] fill between[of=dnn_upper and dnn_lower];
    % --------------------------------------------------------

    % --------------------------------------------------------
    % Noise-resilient
    \addplot [
        color=blue!70!white,          
        dashed,             
        line width=1pt     
        ]   
      coordinates {
      (2\%, 0.08)
      (5\%, 0.08)
      (10\%, 0.08)
      (50\%, 0.08)
      (75\%, 0.08)};
    % Standard deviation region (example values for std)
    \addplot[name path=resilient_upper, draw=none] 
    coordinates {
      (2\%, 0) 
      (5\%, 0) 
      (10\%, 0) 
      (50\%, 0) 
      (75\%, 0)};   
    \addplot[name path=resilient_lower, draw=none] 
    coordinates {
      (2\%, 0) 
      (5\%, 0) 
      (10\%, 0) 
      (50\%, 0) 
      (75\%, 0)};
    % Fill standard deviation region
    \addplot[blue!30!white, opacity=0.5] fill between[of=resilient_upper and resilient_lower];
    % --------------------------------------------------------
          
  \end{axis}
\end{tikzpicture}
    }
    \caption{Kripke--computation}\label{fig:app_kripke1}
  \end{subfigure}  
\hfill
 \begin{subfigure}[t]{0.32\textwidth} 
    \centering
    \resizebox{\columnwidth}{!}{%
        \makeatletter
\newcommand\HUGE{\@setfontsize\Huge{25}{25}}
\newcommand\MID{\@setfontsize\Mid{20}{20}}
\makeatother

\begin{tikzpicture}
  \centering
  \begin{axis}[
        height=8.5cm, width=11cm,
        bar width=1.0cm,
        ymajorgrids, tick align=inside,
        major grid style={draw=gray!20!white},
        enlarge y limits={value=.1,upper},
        ymin=0, ymax=0.6,
        axis x line*=bottom,
        axis y line*=left,
        y axis line style={opacity=0},
        tickwidth=0pt,
        enlarge x limits=true,
        legend style={
            draw=none,
            at={(0.5,-0.3)},
            anchor=center,
            legend columns=-1,
            mark size=6,
            mark options={scale=0.6},
            /tikz/every even column/.append style={column sep=0.5cm}            
        },
        tick label style = {font=\MID}, 
        ylabel={Exponent deviation},
        ylabel style={at={(-0.06,0.5)}},
        ylabel style={font=\MID},
        xlabel={Artificial noise},
        xlabel style={at={(0.5,-0.03)}},
        xlabel style={font=\MID},
        symbolic x coords={
           2\%, 5\%, 10\%, 50\%, 75\%},
       xtick=data,
       nodes near coords={}
       %nodes near coords={
       % \pgfmathprintnumber[precision=0]{\pgfplotspointmeta}
       %}
    ]  

    % --------------------------------------------------------
    % Noise-sensitive
    \addplot [
        only marks, 
        mark=square*, 
        mark size=6,
        color=orange!70!white,  
        ] 
      coordinates {
      (2\%, 0.513) 
      (5\%, 0.486) 
      (10\%, 0.478) 
      (50\%, 0.461) 
      (75\%, 0.436)};
    % --------------------------------------------------------

    % --------------------------------------------------------
    % DNN-based
    \addplot [
        only marks, 
        mark=triangle*, 
        mark size=6,
        color=purple!80!white,
        ] 
      coordinates {
      (2\%, 0.4605555555555555)
        (5\%, 0.3055555555555555)
        (10\%, 0.27777777777777773)
        (50\%, 0.3688888888888889)
        (75\%, 0.4558333333333333)};
        
    % -------------------------------------------------------- 

    % --------------------------------------------------------
    % Noise-resilient
    \addplot [
        only marks, 
        mark=*, 
        mark size=6,
        blue!70!white
        ] 
      coordinates {
      (2\%, 0.111) 
      (5\%, 0.111) 
      (10\%, 0.111) 
      (50\%, 0.111) 
      (75\%, 0.111)}; 
    % --------------------------------------------------------

    % --------------------------------------------------------
    % Noise-sensitive
    \addplot [
        color=orange!70!white,            
        dashed,             
        line width=1pt     
        ]   
      coordinates {
      (2\%, 0.513) 
      (5\%, 0.486) 
      (10\%, 0.478) 
      (50\%, 0.461) 
      (75\%, 0.436)};
    % Standard deviation region (example values for std)
    \addplot[name path=sensitive_upper, draw=none] 
    coordinates {
      (2\%, 0.545) 
      (5\%, 0.528) 
      (10\%, 0.521) 
      (50\%, 0.557) 
      (75\%, 0.549)};   
    \addplot[name path=sensitive_lower, draw=none] 
    coordinates {
      (2\%, 0.481) 
      (5\%, 0.445) 
      (10\%, 0.434) 
      (50\%, 0.365) 
      (75\%, 0.323)};
    % Fill standard deviation region
    \addplot[color=orange!40!white, opacity=0.5] fill between[of=sensitive_upper and sensitive_lower];
    % --------------------------------------------------------

    % --------------------------------------------------------
    % DNN-based
    \addplot [
        color=purple!80!white,         
        dashed,             
        line width=1pt     
        ]   
      coordinates {
      (2\%, 0.4605555555555555)
        (5\%, 0.3055555555555555)
        (10\%, 0.27777777777777773)
        (50\%, 0.3688888888888889)
        (75\%, 0.4558333333333333)};
    % Standard deviation region (example values for std)
    \addplot[name path=dnn_upper, draw=none] 
    coordinates {
      (2\%, 0.5871308149718265)
        (5\%, 0.3055555555555555)
        (10\%, 0.27777777777777773)
        (50\%, 0.4581312416753557)
        (75\%, 0.5624866166372531)}; 
    \addplot[name path=dnn_lower, draw=none] 
    coordinates {
      (2\%, 0.3339802961392845)
        (5\%, 0.3055555555555555)
        (10\%, 0.27777777777777773)
        (50\%, 0.27964653610242207)
        (75\%, 0.34918005002941355)};
    % Fill standard deviation region
    \addplot[color=purple!40!white, opacity=0.5] fill between[of=dnn_upper and dnn_lower];
    % --------------------------------------------------------

    % --------------------------------------------------------
    % Noise-resilient
    \addplot [
        color=blue!70!white,          
        dashed,             
        line width=1pt     
        ]   
      coordinates {
      (2\%, 0.111) 
      (5\%, 0.111) 
      (10\%, 0.111) 
      (50\%, 0.111) 
      (75\%, 0.111)};
    % Standard deviation region (example values for std)
    \addplot[name path=resilient_upper, draw=none] 
    coordinates {
      (2\%, 0) 
      (5\%, 0) 
      (10\%, 0) 
      (50\%, 0) 
      (75\%, 0)};   
    \addplot[name path=resilient_lower, draw=none] 
    coordinates {
      (2\%, 0) 
      (5\%, 0) 
      (10\%, 0) 
      (50\%, 0) 
      (75\%, 0)};
    % Fill standard deviation region
    \addplot[blue!30!white, opacity=0.5] fill between[of=resilient_upper and resilient_lower];
    % --------------------------------------------------------
          
  \end{axis}
\end{tikzpicture}

% -----------------------------------------------
    }
    \caption{Kripke--communication}\label{fig:app_kripke2}
  \end{subfigure} 
\hfill
\begin{subfigure}[t]{0.32\textwidth} 
    \centering
    \resizebox{\columnwidth}{!}{%
        \makeatletter
\newcommand\HUGE{\@setfontsize\Huge{25}{25}}
\newcommand\MID{\@setfontsize\Mid{20}{20}}
\makeatother

\begin{tikzpicture}
  \centering
  \begin{axis}[
        height=8.5cm, width=11cm,
        bar width=1.0cm,
        ymajorgrids, tick align=inside,
        major grid style={draw=gray!20!white},
        enlarge y limits={value=.1,upper},
        ymin=0, ymax=0.6,
        axis x line*=bottom,
        axis y line*=left,
        y axis line style={opacity=0},
        tickwidth=0pt,
        enlarge x limits=true,
        legend style={
            draw=none,
            at={(0.5,-0.3)},
            anchor=center,
            legend columns=-1,
            mark size=6,
            mark options={scale=0.6},
            /tikz/every even column/.append style={column sep=0.5cm}            
        },
        tick label style = {font=\MID}, 
        ylabel={Exponent deviation},
        ylabel style={at={(-0.06,0.5)}},
        ylabel style={font=\MID},
        xlabel={Artificial noise},
        xlabel style={at={(0.5,-0.03)}},
        xlabel style={font=\MID},
        symbolic x coords={
           2\%, 5\%, 10\%, 50\%, 75\%},
       xtick=data,
       nodes near coords={}
       %nodes near coords={
       % \pgfmathprintnumber[precision=0]{\pgfplotspointmeta}
       %}
    ]  

    % --------------------------------------------------------
    % Noise-sensitive
    \addplot [
        only marks, 
        mark=square*, 
        mark size=6,
        color=orange!70!white,   
        ] 
      coordinates {
      (2\%, 0.301) 
      (5\%, 0.294) 
      (10\%, 0.307) 
      (50\%, 0.441) 
      (75\%, 0.496)};
    % --------------------------------------------------------

    % --------------------------------------------------------
    % DNN-based
    \addplot [
        only marks, 
        mark=triangle*, 
        mark size=6,
        color=purple!80!white,
        ] 
      coordinates {
      (2\%, 0.5)
        (5\%, 0.5)
        (10\%, 0.4958333333333333)
        (50\%, 0.4719166666666667)
        (75\%, 0.47083333333333327)};
        
    % -------------------------------------------------------- 

    % --------------------------------------------------------
    % Noise-resilient
    \addplot [
        only marks, 
        mark=*, 
        mark size=6,
        blue!70!white
        ] 
      coordinates {
      (2\%, 0.125) 
      (5\%, 0.125) 
      (10\%, 0.125) 
      (50\%, 0.125) 
      (75\%, 0.125)};
    % --------------------------------------------------------

    % --------------------------------------------------------
    % Noise-sensitive
    \addplot [
        color=orange!70!white,         
        dashed,             
        line width=1pt     
        ]   
      coordinates {
      (2\%, 0.301) 
      (5\%, 0.294) 
      (10\%, 0.307) 
      (50\%, 0.441) 
      (75\%, 0.496)};
    % Standard deviation region (example values for std)
    \addplot[name path=sensitive_upper, draw=none] 
    coordinates {
     (2\%, 0.320)  
     (5\%, 0.349)  
     (10\%, 0.388)  
     (50\%, 0.608)  
     (75\%, 0.704)};   
    \addplot[name path=sensitive_lower, draw=none] 
    coordinates {
     (2\%, 0.283)  
     (5\%, 0.240)  
     (10\%, 0.227)  
     (50\%, 0.276)  
     (75\%, 0.289)};  
    % Fill standard deviation region
    \addplot[color=orange!40!white, opacity=0.5] fill between[of=sensitive_upper and sensitive_lower];
    % --------------------------------------------------------

    % --------------------------------------------------------
    % DNN-based
    \addplot [
        color=purple!80!white,         
        dashed,             
        line width=1pt     
        ]   
      coordinates {
      (2\%, 0.5)
        (5\%, 0.5)
        (10\%, 0.4958333333333333)
        (50\%, 0.4719166666666667)
        (75\%, 0.47083333333333327)};
    % Standard deviation region (example values for std)
    \addplot[name path=dnn_upper, draw=none] 
    coordinates {
      (2\%, 0.5)
        (5\%, 0.5)
        (10\%, 0.520483665762915)
        (50\%, 0.5924631941063976)
        (75\%, 0.6027134346038097)}; 
    \addplot[name path=dnn_lower, draw=none] 
    coordinates {
      (2\%, 0.5)
        (5\%, 0.5)
        (10\%, 0.47118300090375154)
        (50\%, 0.3513701392269359)
        (75\%, 0.3389532320628569)};
    % Fill standard deviation region
    \addplot[color=purple!40!white, opacity=0.5] fill between[of=dnn_upper and dnn_lower];
    % --------------------------------------------------------

    % --------------------------------------------------------
    % Noise-resilient
    \addplot [
        color=blue!70!white,          
        dashed,             
        line width=1pt     
        ]   
      coordinates {
      (2\%, 0.125) 
      (5\%, 0.125) 
      (10\%, 0.125) 
      (50\%, 0.125) 
      (75\%, 0.125)};
    % Standard deviation region (example values for std)
    \addplot[name path=resilient_upper, draw=none] 
    coordinates {
      (2\%, 0.125) 
      (5\%, 0.125) 
      (10\%, 0.125) 
      (50\%, 0.125) 
      (75\%, 0.125)}; 
    \addplot[name path=resilient_lower, draw=none] 
    coordinates {
      (2\%, 0.125) 
      (5\%, 0.125) 
      (10\%, 0.125) 
      (50\%, 0.125) 
      (75\%, 0.125)};
    % Fill standard deviation region
    \addplot[blue!30!white, opacity=0.5] fill between[of=resilient_upper and resilient_lower];
    % --------------------------------------------------------
          
  \end{axis}
\end{tikzpicture}

% -----------------------------------------------
    }
    \caption{RELeARN}\label{fig:app_relearn1}
  \end{subfigure} 
  \medskip      
 \begin{subfigure}[t]{0.32\textwidth} 
    \centering
    \resizebox{\columnwidth}{!}{%
        \makeatletter
\newcommand\HUGE{\@setfontsize\Huge{25}{25}}
\newcommand\MID{\@setfontsize\Mid{20}{20}}
\makeatother

\begin{tikzpicture}
  \centering
  \begin{axis}[
        height=8.5cm, width=11cm,
        bar width=1.0cm,
        ymajorgrids, tick align=inside,
        major grid style={draw=gray!20!white},
        enlarge y limits={value=.1,upper},
        ymin=0, ymax=100,
        axis x line*=bottom,
        axis y line*=left,
        y axis line style={opacity=0},
        tickwidth=0pt,
        enlarge x limits=true,
        legend style={
            draw=none,
            at={(0.5,-0.3)},
            anchor=center,
            legend columns=-1,
            /tikz/every even column/.append style={column sep=0.5cm}            
        },
        %legend to name={legend_noise},
        tick label style = {font=\MID}, 
        ylabel={Relative error (\%)},
        ylabel style={at={(-0.06,0.5)}},
        ylabel style={font=\MID},
        xlabel={Artificial noise},
        xlabel style={at={(0.5,-0.03)}},
        xlabel style={font=\MID},
        symbolic x coords={
           2\%, 5\%, 10\%, 50\%, 75\%},
       xtick=data,
       nodes near coords={}
       %nodes near coords={
       % \pgfmathprintnumber[precision=0]{\pgfplotspointmeta}
       %}
    ]
    
   % --------------------------------------------------------
    % Noise-sensitive
    \addplot [
        only marks, 
        mark=square*, 
        mark size=6,
        color=orange!70!white,  
        ] 
      coordinates {
      (2\%, 29.079) 
      (5\%, 30.425) 
      (10\%, 32.941) 
      (50\%, 59.348) 
      (75\%, 76.984)};
    % --------------------------------------------------------

    % --------------------------------------------------------
    % DNN-based
    \addplot [
        only marks, 
        mark=triangle*, 
        mark size=6,
        color=purple!80!white,
        ] 
      coordinates {
      (2\%, 28.406391827580297)
        (5\%, 26.67353424241337)
        (10\%, 28.116476287898127)
        (50\%, 57.83108811482284)
        (75\%, 82.466384940916)};
    % -------------------------------------------------------- 

    % --------------------------------------------------------
    % Noise-resilient
    \addplot [
        only marks, 
        mark=*, 
        mark size=6,
        blue!70!white
        ] 
      coordinates {
      (2\%, 11.273) 
      (5\%, 10.817) 
      (10\%, 10.086) 
      (50\%, 19.919) 
      (75\%, 35.170) };
    % --------------------------------------------------------
   
    % --------------------------------------------------------
    % Noise-sensitive
    \addplot [
        color=orange!70!white,            
        dashed,             
        line width=1pt     
        ]   
      coordinates {
      (2\%, 29.079) 
      (5\%, 30.425) 
      (10\%, 32.941) 
      (50\%, 59.348) 
      (75\%, 76.984)};
    % Standard deviation region (example values for std)
    \addplot[name path=sensitive_upper, draw=none] 
    coordinates {
      (2\%, 29.192) 
      (5\%, 30.728) 
      (10\%, 33.973) 
      (50\%, 75.442) 
      (75\%, 95.973)};   
    \addplot[name path=sensitive_lower, draw=none] 
    coordinates {
      (2\%, 28.966) 
      (5\%, 30.122) 
      (10\%, 31.909) 
      (50\%, 43.253) 
      (75\%, 58.996)};
    % Fill standard deviation region
    \addplot[color=orange!40!white, opacity=0.5] fill between[of=sensitive_upper and sensitive_lower];
    % --------------------------------------------------------

    % --------------------------------------------------------
    % DNN-based
    \addplot [
        color=purple!80!white,        
        dashed,             
        line width=1pt     
        ]   
      coordinates {
      (2\%, 28.406391827580297)
        (5\%, 26.67353424241337)
        (10\%, 28.116476287898127)
        (50\%, 57.83108811482284)
        (75\%, 82.466384940916)};
    % Standard deviation region (example values for std)
    \addplot[name path=dnn_upper, draw=none] 
    coordinates {
      (2\%, 33.82254027967387)
        (5\%, 33.12362215964888)
        (10\%, 34.91359671098861)
        (50\%, 69.02754119011202)
        (75\%, 96.71734882910643)}; 
    \addplot[name path=dnn_lower, draw=none] 
    coordinates {
      (2\%, 22.99024337548672)
        (5\%, 20.223446325177864)
        (10\%, 21.319355864807648)
        (50\%, 46.63463503953366)
        (75\%, 68.21542105272557)};
    % Fill standard deviation region
    \addplot[color=purple!40!white, opacity=0.5] fill between[of=dnn_upper and dnn_lower];
    % --------------------------------------------------------

    % --------------------------------------------------------
    % Noise-resilient
    \addplot [
        color=blue!70!white,          
        dashed,             
        line width=1pt     
        ]   
      coordinates {
      (2\%, 11.273) 
      (5\%, 10.817) 
      (10\%, 10.086) 
      (50\%, 19.919) 
      (75\%, 35.170) };
    % Standard deviation region (example values for std)
    \addplot[name path=resilient_upper, draw=none] 
    coordinates {
      (2\%, 11.291) 
      (5\%, 10.868) 
      (10\%, 10.190) 
      (50\%, 21.907) 
      (75\%, 38.051)};   
    \addplot[name path=resilient_lower, draw=none] 
    coordinates {
      (2\%, 11.255) 
      (5\%, 10.765) 
      (10\%, 9.982) 
      (50\%, 17.931) 
      (75\%, 32.289)};
    % Fill standard deviation region
    \addplot[blue!30!white, opacity=0.5] fill between[of=resilient_upper and resilient_lower];
    % --------------------------------------------------------
    
  \end{axis}
\end{tikzpicture}

				
    }
    \caption{Kripke--computation}\label{fig:app_re_kripke1}
  \end{subfigure}   
 \begin{subfigure}[t]{0.32\textwidth} 
    \centering
    \resizebox{\columnwidth}{!}{%
        \makeatletter
\newcommand\HUGE{\@setfontsize\Huge{25}{25}}
\newcommand\MID{\@setfontsize\Mid{20}{20}}
\makeatother
\begin{tikzpicture}
  \centering
  \begin{axis}[
        height=8.5cm, width=11cm,
        bar width=1.0cm,
        ymajorgrids, tick align=inside,
        major grid style={draw=gray!20!white},
        enlarge y limits={value=.1,upper},
        ymin=0, ymax=100,
        axis x line*=bottom,
        axis y line*=left,
        y axis line style={opacity=0},
        tickwidth=0pt,
        enlarge x limits=true,
        legend style={
            draw=none,
            at={(0.5,-0.3)},
            anchor=center,
            legend columns=-1,
            /tikz/every even column/.append style={column sep=0.5cm}            
        },
        %legend to name={legend_noise},
        tick label style = {font=\MID}, 
        ylabel={Relative error (\%)},
        ylabel style={at={(-0.06,0.5)}},
        ylabel style={font=\MID},
        xlabel={Artificial noise},
        xlabel style={at={(0.5,-0.03)}},
        xlabel style={font=\MID},
        symbolic x coords={
           2\%, 5\%, 10\%, 50\%, 75\%},
       xtick=data,
       nodes near coords={}
       %nodes near coords={
       % \pgfmathprintnumber[precision=0]{\pgfplotspointmeta}
       %}
    ]
    
  % --------------------------------------------------------
    % Noise-sensitive
    \addplot [
        only marks, 
        mark=square*, 
        mark size=6,
        color=orange!70!white, 
        ] 
      coordinates {
      (2\%, 32.448) 
      (5\%, 34.091) 
      (10\%, 36.384) 
      (50\%, 63.794) 
      (75\%, 84.727)};
    % --------------------------------------------------------

    % --------------------------------------------------------
    % DNN-based
    \addplot [
        only marks, 
        mark=triangle*, 
        mark size=6,
        color=purple!80!white,
        ] 
      coordinates {
      (2\%, 20.047052999333584)
        (5\%, 26.795700080496477)
        (10\%, 21.204707632978753)
        (50\%, 62.69775460509258)
        (75\%, 77.32618537031381)};
    % -------------------------------------------------------- 

    % --------------------------------------------------------
    % Noise-resilient
    \addplot [
        only marks, 
        mark=*, 
        mark size=6,
        blue!70!white
        ] 
      coordinates {
      (2\%, 25.036) 
      (5\%, 23.508) 
      (10\%, 20.894) 
      (50\%, 14.690) 
      (75\%, 20.742)};
    % --------------------------------------------------------

    % --------------------------------------------------------
    % Noise-sensitive
    \addplot [
        color=orange!70!white,           
        dashed,             
        line width=1pt     
        ]   
      coordinates {
      (2\%, 32.448) 
      (5\%, 34.091) 
      (10\%, 36.384) 
      (50\%, 63.794) 
      (75\%, 84.727)};
    % Standard deviation region (example values for std)
    \addplot[name path=sensitive_upper, draw=none] 
    coordinates {
      (2\%, 32.666) 
      (5\%, 35.794) 
      (10\%, 38.220) 
      (50\%, 74.163) 
      (75\%, 99.275)};   
    \addplot[name path=sensitive_lower, draw=none] 
    coordinates {
      (2\%, 32.230) 
      (5\%, 32.388) 
      (10\%, 34.548) 
      (50\%, 53.426) 
      (75\%, 70.178)};
    % Fill standard deviation region
    \addplot[color=orange!40!white, opacity=0.5] fill between[of=sensitive_upper and sensitive_lower];
    % --------------------------------------------------------

    % --------------------------------------------------------
    % DNN-based
    \addplot [
        color=purple!80!white,        
        dashed,             
        line width=1pt     
        ]   
      coordinates {
      (2\%, 20.047052999333584)
        (5\%, 26.795700080496477)
        (10\%, 21.204707632978753)
        (50\%, 62.69775460509258)
        (75\%, 77.32618537031381)};
    % Standard deviation region (example values for std)
    \addplot[name path=dnn_upper, draw=none] 
    coordinates {
      (2\%, 24.15961248921478)
        (5\%, 27.03067670542977)
        (10\%, 21.730142555329955)
        (50\%, 71.19929710959417)
        (75\%, 92.57334452344703)}; 
    \addplot[name path=dnn_lower, draw=none] 
    coordinates {
      (2\%, 15.93449350945239)
        (5\%, 26.560723455563185)
        (10\%, 20.67927271062755)
        (50\%, 54.196212100590984)
        (75\%, 62.079026217180584)};
    % Fill standard deviation region
    \addplot[color=purple!40!white, opacity=0.5] fill between[of=dnn_upper and dnn_lower];
    % --------------------------------------------------------

    % --------------------------------------------------------
    % Noise-resilient
    \addplot [
        color=blue!70!white,          
        dashed,             
        line width=1pt     
        ]   
      coordinates {
      (2\%, 25.036) 
      (5\%, 23.508) 
      (10\%, 20.894) 
      (50\%, 14.690) 
      (75\%, 20.742)};
    % Standard deviation region (example values for std)
    \addplot[name path=resilient_upper, draw=none] 
    coordinates {
      (2\%, 25.159) 
      (5\%, 23.757) 
      (10\%, 21.424) 
      (50\%, 15.929) 
      (75\%, 22.814)};   
    \addplot[name path=resilient_lower, draw=none] 
    coordinates {
      (2\%, 24.914) 
      (5\%, 23.260) 
      (10\%, 20.364) 
      (50\%, 13.452) 
      (75\%, 18.670)};
    % Fill standard deviation region
    \addplot[blue!30!white, opacity=0.5] fill between[of=resilient_upper and resilient_lower];
    % --------------------------------------------------------
    
  \end{axis}
\end{tikzpicture}
    }
    \caption{Kripke--communication}\label{fig:app_re_kripke2}
  \end{subfigure} 
\hfill
 \begin{subfigure}[t]{0.32\textwidth} 
    \centering
    \resizebox{\columnwidth}{!}{%
        \makeatletter
\newcommand\HUGE{\@setfontsize\Huge{25}{25}}
\newcommand\MID{\@setfontsize\Mid{20}{20}}
\makeatother

\begin{tikzpicture}
  \centering
  \begin{axis}[
        height=8.5cm, width=11cm,
        bar width=1.0cm,
        ymajorgrids, tick align=inside,
        major grid style={draw=gray!20!white},
        enlarge y limits={value=.1,upper},
        ymin=0, ymax=100,
        axis x line*=bottom,
        axis y line*=left,
        y axis line style={opacity=0},
        tickwidth=0pt,
        enlarge x limits=true,
        legend style={
            draw=none,
            at={(0.5,-0.3)},
            anchor=center,
            legend columns=-1,
            /tikz/every even column/.append style={column sep=0.5cm}            
        },
        %legend to name={legend_noise},
        tick label style = {font=\MID}, 
        ylabel={Relative error (\%)},
        ylabel style={at={(-0.06,0.5)}},
        ylabel style={font=\MID},
        xlabel={Artificial noise},
        xlabel style={at={(0.5,-0.03)}},
        xlabel style={font=\MID},
        symbolic x coords={
           2\%, 5\%, 10\%, 50\%, 75\%},
       xtick=data,
       nodes near coords={}
    ]
    
   % --------------------------------------------------------
    % Noise-sensitive
    \addplot [
        only marks, 
        mark=square*, 
        mark size=6,
        color=orange!70!white,  
        ] 
      coordinates {
     (2\%, 13.534)  
     (5\%, 15.860)  
     (10\%, 19.657)  
     (50\%, 52.995)  
     (75\%, 70.420)};
    % --------------------------------------------------------

    % --------------------------------------------------------
    % DNN-based
    \addplot [
        only marks, 
        mark=triangle*, 
        mark size=6,
        color=purple!80!white, 
        ] 
      coordinates {
      (2\%, 21.507798343650084)
        (5\%, 24.085985094173594)
        (10\%, 27.85585293624154)
        (50\%, 58.894042326876686)
        (75\%, 78.18593259630207)}; 
    % -------------------------------------------------------- 

    % --------------------------------------------------------
    % Noise-resilient
    \addplot [
        only marks, 
        mark=*, 
        mark size=6,
        blue!70!white
        ] 
      coordinates {
      (2\%, 3.625)  
      (5\%, 3.701)  
      (10\%, 6.107)  
      (50\%, 33.403)  
      (75\%, 50.440)};
    % --------------------------------------------------------

    % --------------------------------------------------------
    % Noise-sensitive
    \addplot [
        color=orange!70!white,           
        dashed,             
        line width=1pt     
        ]   
     coordinates {
     (2\%, 13.534)  
     (5\%, 15.860)  
     (10\%, 19.657)  
     (50\%, 52.995)  
     (75\%, 70.420)};
    % Standard deviation region (example values for std)       
    \addplot[name path=sensitive_upper, draw=none] 
    coordinates {
      (2\%, 13.806)  
     (5\%, 16.496)  
     (10\%, 21.097)  
     (50\%, 60.537)  
     (75\%, 80.671)};
    \addplot[name path=sensitive_lower, draw=none] 
    coordinates {
      (2\%, 13.262)  
      (5\%, 15.224)  
      (10\%, 18.217)  
      (50\%, 45.453)  
      (75\%, 60.169)};
    % Fill standard deviation region
    \addplot[color=orange!40!white, opacity=0.5] fill between[of=sensitive_upper and sensitive_lower];
    % --------------------------------------------------------

    % --------------------------------------------------------
    % DNN-based
    \addplot [
        color=purple!80!white,          
        dashed,             
        line width=1pt     
        ]   
      coordinates {
      (2\%, 21.507798343650084)
        (5\%, 24.085985094173594)
        (10\%, 27.85585293624154)
        (50\%, 58.894042326876686)
        (75\%, 78.18593259630207)};
    % Standard deviation region (example values for std)
    \addplot[name path=dnn_upper, draw=none] 
    coordinates {
      (2\%, 21.716167999245446)
        (5\%, 24.634645045500385)
        (10\%, 29.414485801348988)
        (50\%, 66.49479109209798)
        (75\%, 88.63443167508247)}; 
    \addplot[name path=dnn_lower, draw=none] 
    coordinates {
      (2\%, 21.29942868805472)
        (5\%, 23.537325142846804)
        (10\%, 26.29722007113409)
        (50\%, 51.29329356165539)
        (75\%, 67.73743351752167)};
    % Fill standard deviation region
    \addplot[color=purple!40!white, opacity=0.5] fill between[of=dnn_upper and dnn_lower];
    % --------------------------------------------------------

    % --------------------------------------------------------
    % Noise-resilient
    \addplot [
        color=blue!70!white,          
        dashed,             
        line width=1pt     
        ]   
      coordinates {
      (2\%, 3.625)  
      (5\%, 3.701)  
      (10\%, 6.107)  
      (50\%, 33.403)  
      (75\%, 50.440)};
    % Standard deviation region (example values for std)
    \addplot[name path=resilient_upper, draw=none] 
    coordinates {
     (2\%, 3.601)  
     (5\%, 3.643)  
     (10\%, 5.550)  
     (50\%, 30.147)  
     (75\%, 45.667)};   
    \addplot[name path=resilient_lower, draw=none] 
    coordinates {
     (2\%, 3.648)  
     (5\%, 3.758)  
     (10\%, 6.664)  
     (50\%, 36.658)  
     (75\%, 55.212)};
    % Fill standard deviation region
    \addplot[blue!30!white, opacity=0.5] fill between[of=resilient_upper and resilient_lower];
    % --------------------------------------------------------
    
  \end{axis}
\end{tikzpicture}
    }
    \caption{RELeARN}\label{fig:app_re_relearn1}
  \end{subfigure} 
  \ref{legend_noise}
  \caption{Robustness of performance models under artificial noise. The figure illustrates the mean and standard deviation of our evaluation metrics---exponent deviation and relative error---at test points after introducing varying levels of artificial noise into runtime measurements.}
  %\RemarkAlex{I think the lines between the marks are a bit too thin.}}
  \label{fig:app_noise}
\end{figure*}

\subsubsection{Experimental costs} \label{sec:eval:real:cost}

A drawback of counting basic blocks with \scorep is the extra overhead beyond measuring the default metrics (e.g., runtime in seconds). To quantify this overhead, we measure the execution time of Kripke and RELeARN at different test points $\left( p, G, Z \right) = \left(4096, 160, 20^{3}\right)$ and $\left( p, n \right) = \left(512, 450\right)$. In the default profiling mode, instrumentation introduces execution overheads of $18\%$ and $11\%$ for Kripke and RELeARN, increasing to $59\%$ and $99\%$ with basic-block counting enabled. This overhead is application-dependent and more prominent in our case studies compared to synthetic analysis, because real applications frequently execute numerous low-cost functions. However, the overhead of the basic-block counter affects only execution time, but not the collected basic-block metrics.
 
As outlined in~\cref{sec:eval:method:measurements}, classical performance modeling requires five repetitions per configuration to mitigate noise effects. Our method reduces the number of required repetitions from five to two (one repetition for runtime + one repetition for effort). Notably, the extra overhead applies only to the noise-resilient effort portion. Profiling Kripke using the classical method incurs a total cost of $5.9 \times$ the non-instrumented execution time (i.e., five executions, each with the standard \scorep overhead).  Our method lowers this to $2.8 \times$, comprising one runtime measurement with the standard \scorep overhead and one effort measurement with the \scorep overhead specific to effort profiling. This results in a $52\%$ reduction in experimental cost.  Similarly, for RELeARN, the classical approach requires in total $5.5 \times$ the non-instrumented runtime, while our method achieves $3.1 \times$, corresponding to a $44\%$ reduction. Overall, even with the extra overhead introduced by the noise-resilient effort measurement, our method cuts the total measurement time roughly in half compared to the classical method.

To evaluate model stability under reduced measurement frequency, we vary repetition counts from $1$ to $5$ and test all subsets. \Cref{fig:app_reduction} presents the resulting ED and RE statistics. SWC-based models maintain stable behavior with low variance. In contrast, classic and DNN-based models fluctuate substantially, particularly for Kripke---, which is up to five times more than SWC-based models. For the RE, our approach not only maintains lower variance but also reduces the average error for Kripke–computation and RELeARN. While the DNN-based model shows slightly lower RE for Kripke–communication, its variance is significantly higher. Overall, repeated measurements offer minimal gains for our new method, as a single run is typically sufficient to achieve accuracy. 

\begin{figure*}[tp!]
\centering
  \begin{subfigure}[t]{0.32\textwidth} 
    \centering
    \resizebox{\columnwidth}{!}{%
        \makeatletter
\newcommand\HUGE{\@setfontsize\Huge{25}{25}}
\newcommand\MID{\@setfontsize\Mid{20}{20}}
\makeatother

\begin{tikzpicture}
  \centering
  \begin{axis}[
        height=8.5cm, width=11cm,
        bar width=1.0cm,
        ymajorgrids, tick align=inside,
        major grid style={draw=gray!20!white},
        enlarge y limits={value=.1,upper},
        ymin=0, ymax=1.0,
        axis x line*=bottom,
        axis y line*=left,
        y axis line style={opacity=0},
        tickwidth=0pt,
        enlarge x limits=true,
        legend style={
            draw=none,
            at={(0.5,-0.3)},
            anchor=center,
            legend columns=-1,
            mark size=6,
            mark options={scale=0.6},
            /tikz/every even column/.append style={column sep=0.5cm}            
        },
        legend to name={legend_reduction},
        tick label style = {font=\MID}, 
        ylabel={Exponent deviation},
        ylabel style={at={(-0.06,0.5)}},
        ylabel style={font=\MID},
        xlabel={Number of repetitions},
        xlabel style={at={(0.5,-0.03)}},
        xlabel style={font=\MID},
        symbolic x coords={
           1, 2, 3, 4, 5},
       xtick=data,
       nodes near coords={}
    ]       
    
    % --------------------------------------------------------
    % Noise-sensitive
    \addplot [
        only marks, 
        mark=square*, 
        mark size=6,
        color=orange!70!white, 
        ] 
      coordinates {
      (1, 0.5967) 
      (2, 0.4994) 
      (3, 0.5267) 
      (4, 0.4433) 
      (5, 0.483)};
      \addlegendentry[color=black]{Classic}        
    % --------------------------------------------------------

    % --------------------------------------------------------
    % DNN-based
    \addplot [
        only marks, 
        mark=triangle*, 
        mark size=6,
        purple!80!white
        ] 
      coordinates {
      (1, 0.5277777777777778)
        (2, 0.3111111111111111)
        (3, 0.4111111111111111)
        (4, 0.45555555555555555)
        (5, 0.5833333333333334)};
    \addlegendentry[color=black]{DNN-based}  
    % -------------------------------------------------------- 

    % --------------------------------------------------------
    % Noise-resilient
    \addplot [
        only marks, 
        mark=*, 
        mark size=6,
        blue!70!white
        ] 
      coordinates {
      (1, 0.083) 
      (2, 0.083) 
      (3, 0.083) 
      (4, 0.083) 
      (5, 0.083)};
      \addlegendentry[color=black]{SWC-based} 
    % --------------------------------------------------------

    % --------------------------------------------------------
    % Noise-sensitive
    \addplot [
        orange!70!white,         
        dashed,             
        line width=1pt     
        ]   
      coordinates {
      (1, 0.5967) 
      (2, 0.4994) 
      (3, 0.5267) 
      (4, 0.4433) 
      (5, 0.483)};
    % Standard deviation region (example values for std)
    \addplot[name path=sensitive_upper, draw=none] 
    coordinates {
      (1, 0.9139) 
      (2, 0.7308) 
      (3, 0.6720) 
      (4, 0.4884) 
      (5, 0.483)};  
    \addplot[name path=sensitive_lower, draw=none] 
    coordinates {
      (1, 0.2794) 
      (2, 0.2681) 
      (3, 0.3814) 
      (4, 0.3983) 
      (5, 0.483)};
    % Fill standard deviation region
    \addplot[color=orange!40!white, opacity=0.5] fill between[of=sensitive_upper and sensitive_lower];
    % --------------------------------------------------------

    % --------------------------------------------------------
    % DNN-based
    \addplot [
        purple!80!white,          
        dashed,             
        line width=1pt     
        ]   
      coordinates {
      (1, 0.5277777777777778)
        (2, 0.3111111111111111)
        (3, 0.4111111111111111)
        (4, 0.45555555555555555)
        (5, 0.5833333333333334)};
    % Standard deviation region (example values for std)
    \addplot[name path=dnn_upper, draw=none] 
    coordinates {
      (1, 0.5277777777777778)
        (2, 0.37542131612661234)
        (3, 0.5055555555555555)
        (4, 0.5748383919643549)
        (5, 0.5833333333333334)};
    \addplot[name path=dnn_lower, draw=none] 
    coordinates {
      (1, 0.5277777777777778)
        (2, 0.24680090609560987)
        (3, 0.31666666666666665)
        (4, 0.33627271914675616)
        (5, 0.5833333333333334)};
    % Fill standard deviation region
    \addplot[color=purple!40!white, opacity=0.5] fill between[of=dnn_upper and dnn_lower];
    % --------------------------------------------------------

    % --------------------------------------------------------
    % Noise-resilient
    \addplot [
        color=blue!70!white,          
        dashed,             
        line width=1pt     
        ]   
      coordinates {
      (1, 0.083) 
      (2, 0.083) 
      (3, 0.083) 
      (4, 0.083) 
      (5, 0.083)};
    % Standard deviation region (example values for std)
    \addplot[name path=resilient_upper, draw=none] 
    coordinates {
      (1, 0.083) 
      (2, 0.083) 
      (3, 0.083) 
      (4, 0.083) 
      (5, 0.083)};
    \addplot[name path=resilient_lower, draw=none] 
    coordinates {
      (1, 0.083) 
      (2, 0.083) 
      (3, 0.083) 
      (4, 0.083) 
      (5, 0.083)};
    % Fill standard deviation region
    \addplot[blue!30!white, opacity=0.5] fill between[of=resilient_upper and resilient_lower];
    % --------------------------------------------------------
   
  \end{axis}
\end{tikzpicture}
    }
    \caption{Kripke--computation}\label{fig:reduction_kripke_comp}
  \end{subfigure}  
\hfill
 \begin{subfigure}[t]{0.32\textwidth}  
    \centering
    \resizebox{\columnwidth}{!}{%
        \makeatletter
\newcommand\HUGE{\@setfontsize\Huge{25}{25}}
\newcommand\MID{\@setfontsize\Mid{20}{20}}
\makeatother

\begin{tikzpicture}
  \centering
  \begin{axis}[
        height=8.5cm, width=11cm,
        bar width=1.0cm,
        ymajorgrids, tick align=inside,
        major grid style={draw=gray!20!white},
        enlarge y limits={value=.1,upper},
        ymin=0, ymax=1.0,
        axis x line*=bottom,
        axis y line*=left,
        y axis line style={opacity=0},
        tickwidth=0pt,
        enlarge x limits=true,
        legend style={
            draw=none,
            at={(0.5,-0.3)},
            anchor=center,
            legend columns=-1,
            mark size=6,
            mark options={scale=0.6},
            /tikz/every even column/.append style={column sep=0.5cm}            
        },
        legend to name={legend_noise_bg},
        tick label style = {font=\MID}, 
        ylabel={Exponent deviation},
        ylabel style={at={(-0.06,0.5)}},
        ylabel style={font=\MID},
        xlabel={Number of repetitions},
        xlabel style={at={(0.5,-0.03)}},
        xlabel style={font=\MID},
        symbolic x coords={
           1, 2, 3, 4, 5},
       xtick=data,
       nodes near coords={}
       %nodes near coords={
       % \pgfmathprintnumber[precision=0]{\pgfplotspointmeta}
       %}
    ] 

    % --------------------------------------------------------
    % Noise-sensitive
    \addplot [
        only marks, 
        mark=square*, 
        mark size=6,
        orange!70!white, 
        ] 
      coordinates {
      (1, 0.5100) 
      (2, 0.4383) 
      (3, 0.4139) 
      (4, 0.3800) 
      (5, 0.527)};
    % --------------------------------------------------------

    % --------------------------------------------------------
    % DNN-based
    \addplot [
        only marks, 
        mark=triangle*, 
        mark size=6,
        purple!80!white
        ] 
      coordinates {
      (1, 0.19999999999999998)
        (2, 0.2666666666666666)
        (3, 0.3111111111111111)
        (4, 0.29999999999999993)
        (5, 0.24999999999999997)};
    % -------------------------------------------------------- 

    % --------------------------------------------------------
    % Noise-resilient
    \addplot [
        only marks, 
        mark=*, 
        mark size=6,
        blue!70!white
        ] 
      coordinates {
      (1, 0.111) 
      (2, 0.111) 
      (3, 0.111) 
      (4, 0.111) 
      (5, 0.111)};
    % --------------------------------------------------------

    % --------------------------------------------------------
    % Noise-sensitive
    \addplot [
        color=orange!70!white,           
        dashed,             
        line width=1pt     
        ]   
      coordinates {
      (1, 0.5100) 
      (2, 0.4383) 
      (3, 0.4139) 
      (4, 0.3800) 
      (5, 0.527)};
    % Standard deviation region (example values for std)
    \addplot[name path=sensitive_upper, draw=none] 
    coordinates {
      (1, 0.7479) 
      (2, 0.6457) 
      (3, 0.5582) 
      (4, 0.4946) 
      (5, 0.527)};  
    \addplot[name path=sensitive_lower, draw=none] 
    coordinates {
      (1, 0.2721) 
      (2, 0.2310) 
      (3, 0.2696) 
      (4, 0.2654) 
      (5, 0.527)};
    % Fill standard deviation region
    \addplot[color=orange!40!white, opacity=0.5] fill between[of=sensitive_upper and sensitive_lower];
    % --------------------------------------------------------

    % --------------------------------------------------------
    % DNN-based
    \addplot [
        color=purple!80!white,  
        dashed,             
        line width=1pt     
        ]   
      coordinates {
      (1, 0.19999999999999998)
        (2, 0.2666666666666666)
        (3, 0.3111111111111111)
        (4, 0.29999999999999993)
        (5, 0.24999999999999997)};
    % Standard deviation region (example values for std)
    \addplot[name path=dnn_upper, draw=none] 
    coordinates {
      (1, 0.25931710140017394)
        (2, 0.3806571362704621)
        (3, 0.4156382651234821)
        (4, 0.4088662107903634)
        (5, 0.24999999999999997)};
    \addplot[name path=dnn_lower, draw=none] 
    coordinates {
      (1, 0.14068289859982605)
        (2, 0.1526761970628711)
        (3, 0.2065839570987401)
        (4, 0.19113378920963647)
        (5, 0.24999999999999997)};
    % Fill standard deviation region
    \addplot[color=purple!40!white, opacity=0.5] fill between[of=dnn_upper and dnn_lower];
    % --------------------------------------------------------

    % --------------------------------------------------------
    % Noise-resilient
    \addplot [
        color=blue!70!white,          
        dashed,             
        line width=1pt     
        ]   
      coordinates {
      (1, 0.111) 
      (2, 0.111) 
      (3, 0.111) 
      (4, 0.111) 
      (5, 0.111)};
    % Standard deviation region (example values for std)
    \addplot[name path=resilient_upper, draw=none] 
    coordinates {
      (1, 0.111) 
      (2, 0.111) 
      (3, 0.111) 
      (4, 0.111) 
      (5, 0.111)};
    \addplot[name path=resilient_lower, draw=none] 
    coordinates {
      (1, 0.111) 
      (2, 0.111) 
      (3, 0.111) 
      (4, 0.111) 
      (5, 0.111)};
    % Fill standard deviation region
    \addplot[blue!30!white, opacity=0.5] fill between[of=resilient_upper and resilient_lower];
    % --------------------------------------------------------
   
  \end{axis}
\end{tikzpicture}
    }
    \caption{Kripke--communication}\label{fig:reduction_kripke}
  \end{subfigure} 
\hfill
 \begin{subfigure}[t]{0.32\textwidth}  
    \centering
    \resizebox{\columnwidth}{!}{%
        \makeatletter
\newcommand\HUGE{\@setfontsize\Huge{25}{25}}
\newcommand\MID{\@setfontsize\Mid{20}{20}}
\makeatother

\begin{tikzpicture}
  \centering
  \begin{axis}[
        height=8.5cm, width=11cm,
        bar width=1.0cm,
        ymajorgrids, tick align=inside,
        major grid style={draw=gray!20!white},
        enlarge y limits={value=.1,upper},
        ymin=0, ymax=1.0,
        axis x line*=bottom,
        axis y line*=left,
        y axis line style={opacity=0},
        tickwidth=0pt,
        enlarge x limits=true,
        legend style={
            draw=none,
            at={(0.5,-0.3)},
            anchor=center,
            legend columns=-1,
            mark size=6,
            mark options={scale=0.6},
            /tikz/every even column/.append style={column sep=0.5cm}            
        },
        legend to name={legend_noise_bg},
        tick label style = {font=\MID}, 
        ylabel={Exponent deviation},
        ylabel style={at={(-0.06,0.5)}},
        ylabel style={font=\MID},
        xlabel={Number of repetitions},
        xlabel style={at={(0.5,-0.03)}},
        xlabel style={font=\MID},
        symbolic x coords={
           1, 2, 3, 4, 5},
       xtick=data,
       nodes near coords={}
       %nodes near coords={
       % \pgfmathprintnumber[precision=0]{\pgfplotspointmeta}
       %}
    ]      
    
    % --------------------------------------------------------
    % Traditional (noise-sensitive)
    % --------------------------------------------------------
    \addplot [
        only marks, 
        mark=square*, 
        mark size=6,
        color=orange!70!white,  
        ] 
      coordinates {
      (1, 0.3033) 
      (2, 0.3100) 
      (3, 0.3042) 
      (4, 0.2917) 
      (5, 0.291)};
    % --------------------------------------------------------

    % --------------------------------------------------------
    % DNN-based
    \addplot [
        only marks, 
        mark=triangle*, 
        mark size=6,
        color=purple!80!white,
        ] 
      coordinates {
      (1, 0.5) 
      (2, 0.5) 
      (3, 0.5) 
      (4, 0.5) 
      (5, 0.5)}; 
    % -------------------------------------------------------- 

    % --------------------------------------------------------
    % Noise-resilient
    \addplot [
        only marks, 
        mark=*, 
        mark size=6,
        blue!70!white
        ] 
      coordinates {
      (1, 0.125) 
      (2, 0.125) 
      (3, 0.125) 
      (4, 0.125) 
      (5, 0.125)}; 
    % --------------------------------------------------------

    % --------------------------------------------------------
    % DNN-based
    \addplot [
        color=purple!80!white,        
        dashed,             
        line width=1pt     
        ]   
      coordinates {
      (1, 0.5) 
      (2, 0.5) 
      (3, 0.5) 
      (4, 0.5) 
      (5, 0.5)};
    % Standard deviation region (example values for std)
    \addplot[name path=dnn_upper, draw=none] 
    coordinates {
      (1, 0.5) 
      (2, 0.5) 
      (3, 0.5) 
      (4, 0.5) 
      (5, 0.5)};
    \addplot[name path=dnn_lower, draw=none] 
    coordinates {
      (1, 0.5) 
      (2, 0.5) 
      (3, 0.5) 
      (4, 0.5) 
      (5, 0.5)};
    % Fill standard deviation region
    \addplot[color=purple!40!white, opacity=0.5] fill between[of=dnn_upper and dnn_lower];
    % --------------------------------------------------------
    
    % --------------------------------------------------------
    % Noise-resilient
    \addplot [
        color=blue!70!white,          
        dashed,             
        line width=1pt     
        ]   
      coordinates {
      (1, 0.125) 
      (2, 0.125) 
      (3, 0.125) 
      (4, 0.125) 
      (5, 0.125)};
    % Standard deviation region (example values for std)
    \addplot[name path=resilient_upper, draw=none] 
    coordinates {
      (1, 0.125) 
      (2, 0.125) 
      (3, 0.125) 
      (4, 0.125) 
      (5, 0.125)};
    \addplot[name path=resilient_lower, draw=none] 
    coordinates {
      (1, 0.125) 
      (2, 0.125) 
      (3, 0.125) 
      (4, 0.125) 
      (5, 0.125)};
    % Fill standard deviation region
    \addplot[blue!30!white, opacity=0.5] fill between[of=resilient_upper and resilient_lower];
    % --------------------------------------------------------

    % --------------------------------------------------------
    % Noise-sensitive
    \addplot [
        color=orange!70!white,          
        dashed,             
        line width=1pt     
        ]   
      coordinates {
      (1, 0.3033) 
      (2, 0.3100) 
      (3, 0.3042) 
      (4, 0.2917) 
      (5, 0.291)};
    % Standard deviation region (example values for std)
    \addplot[name path=sensitive_upper, draw=none] 
    coordinates {
      (1, 0.3325) 
      (2, 0.3357) 
      (3, 0.3243) 
      (4, 0.2917) 
      (5, 0.291)};  
    \addplot[name path=sensitive_lower, draw=none] 
    coordinates {
      (1, 0.2741) 
      (2, 0.2843) 
      (3, 0.2840) 
      (4, 0.2917) 
      (5, 0.291)};
    % Fill standard deviation region
    \addplot[color=orange!40!white, opacity=0.5] fill between[of=sensitive_upper and sensitive_lower];
    % --------------------------------------------------------
   
  \end{axis}
\end{tikzpicture}
    }
    \caption{RELeARN}\label{fig:reduction_relearn}
  \end{subfigure} 
 \medskip     
% --------
 \begin{subfigure}[t]{0.32\textwidth} 
    \centering
    \resizebox{\columnwidth}{!}{%
        \makeatletter
\newcommand\HUGE{\@setfontsize\Huge{25}{25}}
\newcommand\MID{\@setfontsize\Mid{20}{20}}
\makeatother

\begin{tikzpicture}
  \centering
  \begin{axis}[
        height=8.5cm, width=11cm,
        bar width=1.0cm,
        ymajorgrids, tick align=inside,
        major grid style={draw=gray!20!white},
        enlarge y limits={value=.1,upper},
        ymin=0, ymax=50,
        axis x line*=bottom,
        axis y line*=left,
        y axis line style={opacity=0},
        tickwidth=0pt,
        enlarge x limits=true,
        legend style={
            draw=none,
            at={(0.5,-0.3)},
            anchor=center,
            legend columns=-1,
            mark size=6,
            mark options={scale=0.6},
            /tikz/every even column/.append style={column sep=0.5cm}          
        },       
        tick label style = {font=\MID}, 
        ylabel={Relative error (\%)},
        ylabel style={at={(-0.06,0.5)}},
        ylabel style={font=\MID},
        xlabel={Number of repetitions},
        xlabel style={at={(0.5,-0.03)}},
        xlabel style={font=\MID},
        symbolic x coords={
           1, 2, 3, 4, 5},
       xtick=data,
       nodes near coords={}
    ]
    % --------------------------------------------------------
    % Noise-sensitive
    \addplot [
        only marks, 
        mark=square*, 
        mark size=6,
        color=orange!70!white, 
        ] 
      coordinates {
      (1, 21.5517) 
      (2, 19.4014) 
      (3, 22.4745) 
      (4, 21.6815) 
      (5, 28.22)};
    % --------------------------------------------------------

    % --------------------------------------------------------
    % DNN-based
    \addplot [
        only marks, 
        mark=triangle*, 
        mark size=6,
        color=purple!80!white,
        ] 
      coordinates {
    (1, 19.29334870355379)
    (2, 17.081337750242522)
    (3, 12.007394240691967)
    (4, 12.520212546546624)
    (5, 12.47351467263112)};
    % -------------------------------------------------------- 

    % --------------------------------------------------------
    % Noise-resilient
    \addplot [
        only marks, 
        mark=*, 
        mark size=6,
        blue!70!white
        ] 
      coordinates {
      (1, 19.4482) 
      (2, 14.8865) 
      (3, 13.7286) 
      (4, 11.8645) 
      (5, 11.57)};
    % --------------------------------------------------------

    % --------------------------------------------------------
    % Noise-sensitive
    \addplot [
        color=orange!70!white,          
        dashed,             
        line width=1pt     
        ]   
      coordinates {
      (1, 21.5517) 
      (2, 19.4014) 
      (3, 22.4745) 
      (4, 21.6815) 
      (5, 28.22)};
    % Standard deviation region (example values for std)
    \addplot[name path=sensitive_upper, draw=none] 
    coordinates {
      (1, 26.4062) 
      (2, 24.6461) 
      (3, 26.9689) 
      (4, 24.9272) 
      (5, 28.22)};  
    \addplot[name path=sensitive_lower, draw=none] 
    coordinates {
      (1, 16.6972) 
      (2, 14.1566) 
      (3, 17.9802) 
      (4, 18.4357) 
      (5, 28.22)};
    % Fill standard deviation region
    \addplot[color=orange!40!white, opacity=0.5] fill between[of=sensitive_upper and sensitive_lower];
    % -------------------------------------------------------- 

    % --------------------------------------------------------
    % DNN-based
    \addplot [
        color=purple!80!white,        
        dashed,             
        line width=1pt     
        ]   
      coordinates {
    (1, 19.29334870355379)
    (2, 17.081337750242522)
    (3, 12.007394240691967)
    (4, 12.520212546546624)
    (5, 12.47351467263112)};
    % Standard deviation region (example values for std)
    \addplot[name path=dnn_upper, draw=none] 
    coordinates {
      (1, 19.29334870355379)
        (2, 25.32375396523149)
        (3, 14.486960061444297)
        (4, 15.030848145042462)
        (5, 12.47351467263112)}; 
    \addplot[name path=dnn_lower, draw=none] 
    coordinates {
      (1, 19.29334870355379)
        (2, 8.838921535253556)
        (3, 9.527828419939638)
        (4, 10.009576948050785)
        (5, 12.47351467263112)};
    % Fill standard deviation region
    \addplot[color=purple!40!white, opacity=0.5] fill between[of=dnn_upper and dnn_lower];
    % --------------------------------------------------------

    % --------------------------------------------------------
    % Noise-resilient
    \addplot [
        color=blue!70!white,          
        dashed,             
        line width=1pt     
        ]   
      coordinates {
      (1, 19.4482) 
      (2, 14.8865) 
      (3, 13.7286) 
      (4, 11.8645) 
      (5, 11.57)};
    % Standard deviation region (example values for std)
    \addplot[name path=resilient_upper, draw=none] 
    coordinates {
      (1, 25.4424) 
      (2, 17.9053) 
      (3, 16.2417) 
      (4, 12.6094) 
      (5, 11.57)};
    \addplot[name path=resilient_lower, draw=none] 
    coordinates {
      (1, 13.4540) 
      (2, 11.8676) 
      (3, 11.2155) 
      (4, 11.1196) 
      (5, 11.57)};
    % Fill standard deviation region
    \addplot[color=blue!30!white, opacity=0.5] fill between[of=resilient_upper and resilient_lower];
    % --------------------------------------------------------
      
  \end{axis}
\end{tikzpicture}

				
    }
    \caption{Kripke--computation}\label{fig:reduction_kripke_comp_re}
  \end{subfigure}  
\hfill
 \begin{subfigure}[t]{0.32\textwidth} 
    \centering
    \resizebox{\columnwidth}{!}{%
        \makeatletter
\newcommand\HUGE{\@setfontsize\Huge{25}{25}}
\newcommand\MID{\@setfontsize\Mid{20}{20}}
\makeatother

\begin{tikzpicture}
  \centering
  \begin{axis}[
        height=8.5cm, width=11cm,
        bar width=1.0cm,
        ymajorgrids, tick align=inside,
        major grid style={draw=gray!20!white},
        enlarge y limits={value=.1,upper},
        ymin=0, ymax=50,
        axis x line*=bottom,
        axis y line*=left,
        y axis line style={opacity=0},
        tickwidth=0pt,
        enlarge x limits=true,
        legend style={
            draw=none,
            at={(0.5,-0.3)},
            anchor=center,
            legend columns=-1,
            /tikz/every even column/.append style={column sep=0.5cm}            
        },     
        tick label style = {font=\MID}, 
        ylabel={Relative error (\%)},
        ylabel style={at={(-0.06,0.5)}},
        ylabel style={font=\MID},
        xlabel={Number of repetitions},
        xlabel style={at={(0.5,-0.03)}},
        xlabel style={font=\MID},
        symbolic x coords={
           1, 2, 3, 4, 5},
       xtick=data,
       nodes near coords={}
       %nodes near coords={
       % \pgfmathprintnumber[precision=0]{\pgfplotspointmeta}
       %}
    ]

    % --------------------------------------------------------
    % Noise-sensitive
    \addplot [
        only marks, 
        mark=square*, 
        mark size=6,
        color=orange!70!white, 
        ] 
      coordinates {
      (1, 32.0078) 
      (2, 30.9857) 
      (3, 31.8364) 
      (4, 32.2135) 
      (5, 31.49)};
    % --------------------------------------------------------

    % --------------------------------------------------------
    % DNN-based
    \addplot [
        only marks, 
        mark=triangle*, 
        mark size=6,
        color=purple!80!white,
        ] 
       coordinates {
      (1, 20.39396216943276)
        (2, 16.65147666862756)
        (3, 21.101847734815504)
        (4, 21.508505195514438)
        (5, 39.96264611214862)};
    % -------------------------------------------------------- 

    % --------------------------------------------------------
    % Noise-resilient
    \addplot [
        only marks, 
        mark=*, 
        mark size=6,
        blue!70!white
        ] 
      coordinates {
      (1, 35.9022) 
      (2, 30.5889) 
      (3, 27.9424) 
      (4, 26.6209)      
      (5, 26.05)};
    % --------------------------------------------------------

    % --------------------------------------------------------
    % Noise-sensitive
    \addplot [
        color=orange!70!white,         
        dashed,             
        line width=1pt     
        ]   
      coordinates {
      (1, 32.0078) 
      (2, 30.9857) 
      (3, 31.8364) 
      (4, 32.2135) 
      (5, 31.49)};
    % Standard deviation region (example values for std)
    \addplot[name path=sensitive_upper, draw=none] 
    coordinates {
      (1, 42.8622) 
      (2, 36.9228) 
      (3, 36.1176) 
      (4, 36.2855) 
      (5, 31.49)};  
    \addplot[name path=sensitive_lower, draw=none] 
    coordinates {
      (1, 21.1534) 
      (2, 25.0486) 
      (3, 27.5553) 
      (4, 28.1416) 
      (5, 31.49)};
    % Fill standard deviation region
    \addplot[color=orange!40!white, opacity=0.5] fill between[of=sensitive_upper and sensitive_lower];
    % -------------------------------------------------------- 

    % --------------------------------------------------------
    % DNN-based 
    \addplot [
        color=purple!80!white,        
        dashed,             
        line width=1pt     
        ]   
      coordinates {
      (1, 20.39396216943276)
        (2, 16.65147666862756)
        (3, 21.101847734815504)
        (4, 21.508505195514438)
        (5, 39.96264611214862)};
    % Standard deviation region (example values for std)
    \addplot[name path=dnn_upper, draw=none] 
    coordinates {
      (1, 30.816222543419613)
        (2, 20.184582264889038)
        (3, 26.007712113310024)
        (4, 26.096693801579335)
        (5, 39.96264611214862)}; 
    \addplot[name path=dnn_lower, draw=none] 
    coordinates {
      (1, 9.971701795445911)
        (2, 13.118371072366081)
        (3, 16.195983356320983)
        (4, 16.92031658944954)
        (5, 39.96264611214862)};
    % Fill standard deviation region
    \addplot[color=purple!40!white, opacity=0.5] fill between[of=dnn_upper and dnn_lower];
    % --------------------------------------------------------

    % --------------------------------------------------------
    % Noise-resilient
    \addplot [
        color=blue!70!white,          
        dashed,             
        line width=1pt     
        ]   
      coordinates {
      (1, 35.9022) 
      (2, 30.5889) 
      (3, 27.9424) 
      (4, 26.6209)      
      (5, 26.05)};
    % Standard deviation region (example values for std)
    \addplot[name path=resilient_upper, draw=none] 
    coordinates {
      (1, 42.4600) 
      (2, 34.6169) 
      (3, 30.7390) 
      (4, 28.6356) 
      (5, 26.05)};
    \addplot[name path=resilient_lower, draw=none] 
    coordinates {
      (1, 29.3445) 
      (2, 26.5608)
      (3, 25.1459) 
      (4, 24.6062)
      (5, 26.05)};
    % Fill standard deviation region
    \addplot[blue!30!white, opacity=0.5] fill between[of=resilient_upper and resilient_lower];
    % --------------------------------------------------------
      
  \end{axis}
\end{tikzpicture}
				
				
    }
    \caption{Kripke - communication}\label{fig:reduction_kripke_re}
  \end{subfigure} 
\hfill
 \begin{subfigure}[t]{0.32\textwidth}  
    \centering
    \resizebox{\columnwidth}{!}{%
        \makeatletter
\newcommand\HUGE{\@setfontsize\Huge{25}{25}}
\newcommand\MID{\@setfontsize\Mid{20}{20}}
\makeatother

\begin{tikzpicture}
  \centering
  \begin{axis}[
        height=8.5cm, width=11cm, 
        bar width=1.0cm,
        ymajorgrids, tick align=inside,
        major grid style={draw=gray!20!white},
        enlarge y limits={value=.1,upper},
        ymin=0, ymax=50,
        axis x line*=bottom,
        axis y line*=left,
        y axis line style={opacity=0},
        tickwidth=0pt,
        enlarge x limits=true,
        legend style={
            draw=none,
            at={(0.5,-0.3)},
            anchor=center,
            legend columns=-1,
            /tikz/every even column/.append style={column sep=0.5cm}            
        },      
        tick label style = {font=\MID}, 
        ylabel={Relative error (\%)},
        ylabel style={at={(-0.06,0.5)}},
        ylabel style={font=\MID},
        xlabel={Number of repetitions},
        xlabel style={at={(0.5,-0.03)}},
        xlabel style={font=\MID},
        symbolic x coords={
           1, 2, 3, 4, 5},
       xtick=data,
       nodes near coords={}
       %nodes near coords={
       % \pgfmathprintnumber[precision=0]{\pgfplotspointmeta}
       %}
    ]
    % --------------------------------------------------------
    % Noise-sensitive
    \addplot [
        only marks, 
        mark=square*, 
        mark size=6,
        color=orange!70!white,  
        ] 
      coordinates {
      (1, 11.9708) 
      (2, 11.9865) 
      (3, 11.9987) 
      (4, 11.9941) 
      (5, 11.998)};
    % --------------------------------------------------------

    % --------------------------------------------------------
    % DNN-based
    \addplot [
        only marks, 
        mark=triangle*, 
        mark size=6,
        color=purple!80!white,
        ] 
      coordinates {
      (1, 15.24148171268383)
        (2, 15.220150741488354)
        (3, 15.20788120178556)
        (4, 15.196288161360695)
        (5, 15.361345619754099)};
    % --------------------------------------------------------

    % --------------------------------------------------------
    % Noise-resilient
    \addplot [
        only marks, 
        mark=*, 
        mark size=6,
        blue!70!white
        ] 
      coordinates {
      (1, 3.5786) 
      (2, 3.5740) 
      (3, 3.5757) 
      (4, 3.5767) 
      (5, 3.577)};
    % --------------------------------------------------------
    
    % --------------------------------------------------------
    % Noise-sensitive
    \addplot [
        color=orange!70!white,           
        dashed,             
        line width=1pt     
        ]   
      coordinates {
      (1, 11.9708) 
      (2, 11.9865) 
      (3, 11.9987) 
      (4, 11.9941) 
      (5, 11.998)};
    % Standard deviation region (example values for std)
    \addplot[name path=sensitive_upper, draw=none] 
    coordinates {
      (1, 12.4859) 
      (2, 12.3157) 
      (3, 12.2417) 
      (4, 12.0916) 
      (5, 11.998)};  
    \addplot[name path=sensitive_lower, draw=none] 
    coordinates {
      (1, 11.4557) 
      (2, 11.6574) 
      (3, 11.7556) 
      (4, 11.8966) 
      (5, 11.998)};
    % Fill standard deviation region
    \addplot[color=orange!40!white, opacity=0.5] fill between[of=sensitive_upper and sensitive_lower];
    % --------------------------------------------------------

    % --------------------------------------------------------
    % DNN
    \addplot [
        color=purple!80!white,          
        dashed,             
        line width=1pt     
        ]   
      coordinates {
      (1, 15.24148171268383)
        (2, 15.220150741488354)
        (3, 15.20788120178556)
        (4, 15.196288161360695)
        (5, 15.361345619754099)};
    % Standard deviation region (example values for std)
    \addplot[name path=dnn_upper, draw=none] 
    coordinates {
      (1, 15.584769944495287)
        (2, 15.42927043642412)
        (3, 15.410469756344561)
        (4, 15.317779020905839)
        (5, 15.361345619754099)}; 
    \addplot[name path=dnn_lower, draw=none] 
    coordinates {
      (1, 14.898193480872374)
        (2, 15.011031046552587)
        (3, 15.005292647226558)
        (4, 15.07479730181555)
        (5, 15.361345619754099)};
    % Fill standard deviation region
    \addplot[color=purple!40!white, opacity=0.5] fill between[of=dnn_upper and dnn_lower];
    % --------------------------------------------------------

     % --------------------------------------------------------
    % Noise-resilient
    \addplot [
        color=blue!70!white,          
        dashed,             
        line width=1pt     
        ]   
      coordinates {
      (1, 3.5786) 
      (2, 3.5740) 
      (3, 3.5757) 
      (4, 3.5767) 
      (5, 3.577)};
    % Standard deviation region (example values for std)
    \addplot[name path=resilient_upper, draw=none] 
    coordinates {
      (1, 3.6439) 
      (2, 3.6149) 
      (3, 3.6085) 
      (4, 3.6011) 
      (5, 3.577)};
    \addplot[name path=resilient_lower, draw=none] 
    coordinates {
      (1, 3.5133) 
      (2, 3.5331) 
      (3, 3.5429) 
      (4, 3.5522) 
      (5, 3.577)};
    % Fill standard deviation region
    \addplot[blue!30!white, opacity=0.5] fill between[of=resilient_upper and resilient_lower];
    % --------------------------------------------------------
    
  \end{axis}
\end{tikzpicture}
    }
    \caption{RELeARN}\label{fig:reduction_relearn_re}
  \end{subfigure} 
  \ref{legend_reduction}
  \caption{The impact of repeated time measurements on the accuracy of the models. The figure illustrates the average and the standard deviation of our evaluation metrics, the exponent deviation between theoretical and empirical models, and the relative error between prediction and measured value at the test points.}
  \label{fig:app_reduction}
\end{figure*}

\section{Related Work} \label{sec:r_work}

In HPC systems, noise can have different sources at various levels of the software stack. Several research papers have been dedicated to analyzing and examining these sources~\cite{extranoise,ritter2022conquering}. One strategy to reduce the effect of noise in performance analysis is to conduct repeated measurements. Yet, Shah et al.~\cite{shah2018estimating} pointed out that noise may still affect the fastest run even when multiple measurements are taken. Exploiting the iterative nature of most applications, they filter noise-induced delays by clustering the code into segments. Repeated executions of segments within the same class should demonstrate behavioral consistency; any discrepancies may be attributed to noise interference.

A range of performance modeling techniques with varying levels of automation has been introduced~\cite{Tallent_ea:2014,vetter-aspen-2012, HammerHEW15}. Numerous state-of-the-art tools support empirical performance modeling, which derives performance models from measurements~\cite{Goldsmith_ea:2007, Carrington_ea:2006, grebhahn2017performance}. 
Two main directions have been followed to tackle noise: either to increase the performance modeling process's robustness or make the measurements more noise-resilient. 
As an example of the former, Duplyakin et al.~\cite{duplyakin2016active} introduced Gaussian process regression (GPR) to the field of performance modeling. By selecting a higher absolute value for the noise value hyperparameter, GPR can more effectively handle noisy data, mitigating the risk of overfitting. However, this adjustment may result in the generation of ``smoother'' models that might not capture all the nuances presented in the data.

On the opposite side, noise-resilient performance measurements have been used to enhance the comprehension of code performance and scaling behaviors. 
In this context, Ritter et al. investigated the noise resilience of hardware counters~\cite{ritter2022conquering}, showing that the majority of counters can be affected by noise. Since our approach relies on LLVM-supported software counters and message transfer sizes, the measurements we obtain remain unaffected by noise interference. Several studies also utilize LLVM-supported software counters~\cite{hao2019profpred, yang2021performance}. Hao et al.~\cite{hao2019multi} use software-counter values from a serial program as input for machine learning algorithms to estimate the code's runtime in a parallel configuration. Running alongside, Zhang et al.~\cite{zhang2017predicting} model communication performance using theoretical performance models in conjunction with neural networks, where the computation is derived from LLVM counters and average instruction runtime. Unlike such AI-based tools, our method does not require any training.
In addition, these studies do not investigate the model's resilience under the impact of noise.

Several approaches rest on hardware counters to build and enhance performance models~\cite{7530001,Nagasaka_Maruyama_Nukada_Endo_Matsuoka_2010,ding2019using,TSAFACKCHETSA2014287}. 
Often, the purpose of these models is to estimate the energy consumption of the application~\cite{Nagasaka_Maruyama_Nukada_Endo_Matsuoka_2010,TSAFACKCHETSA2014287,6787326,9154439}. Nagasaka et al.~\cite{Nagasaka_Maruyama_Nukada_Endo_Matsuoka_2010} use GPU performance counters to estimate the power consumption of a GPU. In contrast, Chetsa et al.~\cite{TSAFACKCHETSA2014287} use performance counters to predict the energy consumption of HPC applications. Unlike these studies, we do not focus on a particular resource. Moreover, because of their apparent noise resilience, we use software rather than hardware counters to enhance our performance models. Ding et al.~\cite{Ding_Lee_Xue_Zheng_2020} present an automatic performance modeling tool that uses hardware counter-assisted profiling to identify 
kernels that consume large run time proportions and non-scalable ones. 
In an earlier work, Ding et al.~\cite{ding2019using} proposed a method for performance modeling using hardware counters, in which computation is quantified using a collection of different logical counters (e.g., CPU cycles waiting for memory and executed instructions), and the size of transferred messages defines communication. To estimate the total runtime, the authors analytically model the influence of various measured resources (i.e., using hardware counters) and, subsequently, use the runtime to determine the coefficients of the equation. 
Compared to our approach, they differ in several aspects. The authors introduce an analytical model that reflects the influence of the problem size and the number of processors only to a limited extent, which is why they still require the user to provide domain knowledge to capture the scaling behavior. In contrast, we generate our models fully automatically, relying on the relative growth of our counters. Furthermore, they do not aim to address the noise problem; they only refer to the necessity of repeated time measurements to increase accuracy, an effort we try to reduce. 

The variance between anticipated and observed performance in MPI collectives has also been examined before. Hunold et al.~\cite{hunold2016automatic} showed how implementations of MPI functions can violate performance guidelines. In addition,  Hunold and Carpen-Amarie~\cite{hunold2018autotuning} empirically identify deviations of MPI collectives and performance guidelines (e.g., executing allreduce operations should not be slower than reduce + broadcast). In the case of deviations, they propose replacing them with the faster guideline implementation. In contrast to them, we highlight deviations from expectations to expose waiting time or the influence of noise.

%\hline

%\RemarkFelix{Here is some additional material from the proposal you might want to consider. I think the work is not closely related, but rather on a more general level of empirical performance modeling. It covers potential reviewers who might expect to be cited.}

%To ease the burden, performance modeling techniques with varying degrees of automation have been introduced~\cite{Tallent_ea:2014,vetter-aspen-2012, HammerHEW15}. Many state-of-the-art tools support empirical performance modeling, a method that derives performance models from measurements~\cite{Goldsmith_ea:2007, Carrington_ea:2006, grebhahn2017performance}, Recently, Duplyakin et al.~\cite{duplyakin2016active} applied Gaussian process regression (GPR) to the field of performance modeling. Here, selecting a higher absolute value for the noise value hyper-parameter can help GPR to better cope with noisy data, reducing the risk of overfitting but at the cost of generating "smoother" models that do not necessarily capture all behaviors present in the data. 

\section{Discussion and conclusion} \label{sec:conclusion}

Our method accurately captures the computational effort of an application in close alignment with its theoretical performance model. In our synthetic evaluation, we obtain models with scaling behavior identical to the theoretical model, while the classic models deviate. With real-world applications, our approach outperforms the existing methods in five out of eight scenarios.

When systems are exposed to minimal noise, classic regression-based and DNN-based modeling yield reasonably accurate results. However, the run-to-run variation on HPC systems can reach up to 70\%~\cite{chunduri2017run}. Once noise exceeds a threshold of approximately 10\%, these models show significant error rates, while our noise-resilient priors substantially improve model robustness. Under artificial noise, our models maintained stable error rates for computation and communication (35\% and 60\%) in the synthetic evaluation, whereas classic methods deteriorated markedly (55\% and 127\%). With real-world applications under noisy conditions, our prediction error remained up to four times lower (20\% vs. 84\%). Notably, even under extreme noise, our models approximate the theoretical asymptotic complexity more closely than purely time-based models, which tend to deviate significantly in proportion to the noise intensity. 

We reduce, if not eliminate, the need for multiple time measurements, reducing experimental costs by roughly half compared to classic regression-based modeling. Regarding predictive power, our SWC-based models achieve lower relative errors than the state of the art for the synthetic evaluation: 35\% vs. 45\% for computation and 60\% vs. 91\% for communication. With realistic applications, our approach consistently demonstrates superior accuracy, with existing methods showing up to ten times higher errors (e.g., 2.7\% vs. 28\%). 

Supported by the improved performance models, users can now better predict the performance behavior of their codes on large-scale systems. The asymptotic complexity reveals general scaling bottlenecks, while the alignment of the model with actual time measurements allows estimating the execution time on previously unseen configurations with higher accuracy. Because the core idea makes little to no assumptions about the underlying system or application, we expect wide-ranging generalization beyond our study’s testbeds and code types. 
Beyond finding scaling bottlenecks, comparing runtime models with our new SWC-based models, which capture only actual communication effort, supports the diagnosis of waiting time. 
As a next step, we plan to refine our method using selected hardware counters, expand our support for MPI by including further operations, and cover OpenMP or GPU parallelism---in addition to MPI.
If accepted, we will make all software artifacts, including the \scorep extension, the noise-resilient version of \extrap, and the benchmark generator, publicly available.

\section*{Acknowledgments}

This work was funded by the Deutsche Forschungsgemeinschaft (DFG, German Research Foundation) – Project No. 449683531 (ExtraNoise). The authors gratefully acknowledge the German Federal Ministry of Education and Research (BMBF) and the Hessian Ministry of Science and Research, Art and Culture (HMWK) for supporting this work as part of the NHR funding. Also, this work is supported by the following grant agreements: SwissTwins (funded by the swiss State Secreteriat for Education, Research and Innnovation), ERC PSAP (grant agreement No 101002047). Furthermore, the authors are thankful for the computing time provided to them on the high-performance computer Lichtenberg II at TU Darmstadt, funded by the German Federal Ministry of Education and Research (BMBF) and the State of Hesse. The authors also gratefully appreciate the Jureca-DC supercomputer at the J\"ulich Supercomputing Centre. 
ChatGPT (version GPT-4-turbo, OpenAI) and Grammarly have been used in this work for language editing and grammar enhancements, to improve the readability of code snippets, and to enhance the presentation of data. No technical content generated by AI technology has been presented as our work.

%\bibliographystyle{IEEEtran}
%\bibliography{references}


\end{document}